\pdfoutput=1
\documentclass[prb,reprint,groupedaddress,floatfix,citeautoscript]{revtex4-2}
\usepackage[english]{babel}

\makeatletter
\def\bbl@set@language#1{%
    \edef\languagename{%
        \ifnum\escapechar=\expandafter`\string#1\@empty
        \else\string#1\@empty\fi}%
    \@ifundefined{babel@language@alias@\languagename}{}{%
        \edef\languagename{\@nameuse{babel@language@alias@\languagename}}%
    }%
    \select@language{\languagename}%
    \expandafter\ifx\csname date\languagename\endcsname\relax\else
    \if@filesw
    \protected@write\@auxout{}{\string\select@language{\languagename}}%
    \bbl@for\bbl@tempa\BabelContentsFiles{%
        \addtocontents{\bbl@tempa}{\xstring\select@language{\languagename}}}%
    \bbl@usehooks{write}{}%
    \fi
    \fi}
\newcommand{\DeclareBabelLanguageAlias}[2]{%
    \global\@namedef{babel@language@alias@#1}{#2}%
}
\makeatother
\DeclareBabelLanguageAlias{en}{english}

\usepackage{hyperref}
\hypersetup{
    colorlinks=true,
    allcolors=red,
    citecolor=blue,
    urlcolor=blue
}
\usepackage{amsmath,amssymb}
\usepackage{mathtools}
\usepackage{physics,bm}
\usepackage{siunitx}
\usepackage{standalone}
\usepackage[version=4]{mhchem}
\usepackage{ifthen}
\usepackage{booktabs,nicematrix,enumitem}
\usepackage{easyReview}

\newcommand{\RN}[1]{%
    \textup{\uppercase\expandafter{\romannumeral#1}}%
}
\newcommand{\Figref}[2][]{\hyperref[#2]{\ref*{#2}\ifthenelse{\equal{#1}{}}{}{\thinspace#1}}}

\begin{document}
%\preprint{}

\title{Anisotropic surface polaritons at isotropic--uniaxial interface: an exact algebraic solution}

\author{K. Yu. Golenitskii}
\email[]{golenitski.k@mail.ioffe.ru}
%\thanks{}
\affiliation{Ioffe Institute, 194021 St.~Petersburg, Russia}
\affiliation{Skolkovo Institute of Science and Technology, 121205 Moscow, Russia}

\date{\today}

\begin{abstract}
Surface polaritons in an anisotropic media posses a strong dependence of the wavevector on the propagation direction, which is called the isofrequency contour.
This can lead to the fact that polariton propagation is possible only in a limited range of angles in the boundary plane.
Notable examples are Dyakonov surface waves at the boundary of two dielectrics and hyperbolic plasmons in a hyperbolic metamaterial.
Exact closed-form solutions of the polariton dispersion equation are known only in special cases: in a weakly anisotropic medium, and in an arbitrary medium for highly symmetric directions of polariton propagation.
This work provides an universal exact solution in algebraic form for surface polariton at the interface of an arbitrary isotropic and an uniaxial media for the case of the optic axis parallel to the boundary.
As an example, it is used to analyze the shapes of isofrequency contours of surface polaritons.
The work brings together previously scattered results of studies on surface polaritons of various types in uniaxial media.
In addition to the cases already considered in the literature, a solution for surface polaritons at the boundary of a isotropic metal and a Type I hyperbolic medium is found.
The case of ``elliptic'' polaritons at the boundary of an anisotropic metal-like medium is apparently analyzed here for the first time.
%The aim of the work is to summarize the known results of theoretical researches on surface polaritons in this case.
\end{abstract}
%\keywords{}

\maketitle

% References should be done using the~\cite, \ref, and \label commands
\section{Introduction}
Surface waves have attracted the attention of researchers for many years.
Specifically, in physics, active research continues on surface electromagnetic waves in a variety of media, both natural and artificial.
In the literature they are also called surface polaritons emphasizing the connection with a dispersive medium.
Due to edge cases it is not always possible to clearly determine when a surface wave becomes a surface polariton.
Therefore, both terms are used equally in the text, often implying the same meaning.

As we know, the basic properties of waves, such as polarization and speed, in an anisotropic medium depend on the direction of propagation.
In particular, the range of possible propagation directions in the boundary plane becomes an important characteristic of surface waves.
This is also called angular existence domain (AED).
In the simple case AED always equal \( 2\pi \) when boundary between materials is isotropic.
For example, surface plasmon polaritons (SPPs) at the boundary of metal and dielectric, if it exists, can propagate in any direction.
In this work we consider the case of anisotropic boundary, and AED in some cases may be less than \( 2\pi \).
An illustrative example is Dyakonov surface waves (DSWs)~\cite{Dyakonov1988NewType} that are propagating along the interface between two transparent media at least one of them is birefringent.
The characteristic AED for DSWs in highly anisotropic natural minerals (\ce{Hg2Cl2}, \ce{TiO2}, \ce{YVO4}) in the optical spectrum is only a few degrees.
AED for surface polaritons may depend on the frequency due to the dispersion of dielectric constants.
All other properties of surface waves, such as the localization length, propagation length, frequency dispersion, and others, may also be highly direction dependent.

\paragraph*{Early studies.}
Surface polartions in anisotropic medium have been studied for a long time.
Early studies examined plasmonic waves and others that have existence conditions similar to SPP (a negative dielectric constant): surface magnetoplasmons in a magnetoactive plasma~\cite{Agranovich1965SurfaceExcitons,Pahkomov1967SurfacePotential,Hussein1969StudyElectromagnetic,Abdel-Shahid1970StudyPotential}, in a semiconductor~\cite{Wallis1974TheorySurface}; surface phonon polaritons (SPhPs)~\cite{Borstel1973InterfaceSurface,Falge1974DispersionGeneralized,Borstel1974SurfaceBulk,Puchkovskaya1978CrystalOptics} and surface phonon plasmons~\cite{Gurevich1975SurfacePlasmonpolaritons,Tarkhanian1975TheorySurface}; surface polaritons in a resonant medium~\cite{Agranovich1965TheoryLongwavelength,Agranovich1975CrystalOptics}, in an uniaxial antiferromagnets~\cite{Camley1982SurfacePolaritons}, and in a general uniaxial medium~\cite{Lyubimov1972SurfaceElectromagnetic}.
See also reviews~\cite{Burstein1974SurfacePolaritons,Mirlin1982SurfacePhonon,*[{The chapter of the book }]Agranovich1982SurfacePolaritons,Kushwaha2001PlasmonsMagnetoplasmons}.

It is well known that SPP in isotropic media is a purely transverse magnetic wave (TM wave).
At first we can roughly take anisotropy of the medium into account as a perturbation leading to small changes on the wave polarization~\cite{Hussein1969StudyElectromagnetic,Hartstein1973SurfacePolaritons}.
The conditions for the existence of a SPP with such a rough approach remain almost unchanged.
Therefore, it was assumed that the negativity of at least one of the principal components of dielectric tensor is the necessary condition for surface polariton existence in a homogeneous \footnote{
    An inhomogeneous ``dielectric'', such as periodic stratified medium or photonic crystal, supports propagation of surface waves due to the existence of stop bands. Now we call them surface electromagnetic Tamm states (Tamm waves) or surface Bloch waves. See \textcite{Yeh1977ElectromagneticPropagation,Dyakov2012SurfaceStates}.
    } medium.
As we now know, this idea turned out to be wrong.

\paragraph*{Dyakonov surface waves.}
One of the earliest works mentioning the existence of surface polariton in anisotropic dielectric is~\cite{Marchevskii1984SingularElectromagnetic}.
It predicts the existence of singular surface polariton, also called exceptional surface wave or surface Voigt wave, in two cases.
One of them is interpreted as the case of two dielectric media.
But the existence condition is requiring extremely strong anisotropy of uniaxial medium \footnote{
    Following our notation necessary conditions: \( 0 < \varepsilon_\perp < \varepsilon_0 \) and \( \varepsilon_\parallel\ > \varepsilon_\perp + 2 \varepsilon_0 \) }, and the authors did not pay much attention to this.
To the best of the author's knowledge, a detailed analysis of the existence of a surface waves in a homogeneous anisotropic dielectric was carried out for the first time in~\cite{Dyakonov1988NewType} by M.\,I.~Dyakonov, in the case of the optic axis parallel to the interface.
Later it was generalized to an arbitrary oriented optic axis in~\cite{Walker1998SurfaceMode,Furs1999GeneralExistence,Alshits2002DispersionlessSurface}.
Now we call these Dyakonov surface waves (DSWs).
Obtained existence condition in~\cite{Dyakonov1988NewType} is less strict and in the strong anisotropy limit includes previous given in~\cite{Marchevskii1984SingularElectromagnetic}.
This statement was noted much later~\cite{Zhou2021SingularExistence}.
DSWs possess two remarkable properties.
Unlike SPPs, the DSWs usually \footnote{
    It may be wrong in the case of two anisotropic media. For example, For example, Dyakonov surface waves at the interface of identical positive uniaxial crystals with crossed optic axes propagate along the bisectors~\cite{Averkiev1990ElectromagneticWaves}.} do not propagate in the most symmetrical directions: along the optic axis or orthogonal to it.
This is due to the second property that the DSW is a hybrid wave mixing two polarizations.
Thus, birefringence of the medium is the crucial property required for DSW.
Other properties and features are described in more detail in the review \cite{Takayama2008DyakonovSurface}.

\paragraph*{Surface hyperbolic plasmons, dyakonov plasmons and others.}
A new round of research on surface polaritons in anisotropic media is associated with metamaterials and structures related to them.
A metamaterial in a certain frequency range may have significantly stronger anisotropy and low losses than natural minerals.
For example, the AED of DSWs in a 2D photonic crystal based on \ce{Si} and \ce{SiO2}~\cite{Artigas2005DyakonovSurface} may be a couple of times wider, especially for optical spectrum.
Hyperbolic media are one of the types of metamaterials in which Dyakonov-type surface waves continue to be actively studied~\cite{Jacob2008OpticalHyperspace,Zapata-Rodriguez2013EngineeredSurface,Takayama2017PhotonicSurface,Takayama2017MidinfraredSurface,Davidovich2020DyakonovPlasmon,Passler2022HyperbolicShear,Moradi2023ThermallyTunable}.
They are named \textit{hyperbolic} due to the shape of isofrequency surfaces for a plane wave, which is a one- or two-sheet hyperboloid~\cite{Poddubny2013HyperbolicMetamaterials}.
This is possible when the components of the dielectric tensor in such a medium have different signs, or in other words, it has an indefinite dielectric tensor.
Despite the fact that such media have been studied for a long time, the term \textit{hyperbolic medium} has been established relatively recently.
A notable example mentioned above is light--optical phonon coupling in an anisotropic crystal that may lead to hyperbolic behavior in some frequency ranges around the frequencies of optical phonons~\cite{Passler2022HyperbolicShear}.
After Dyakonov's paper, already known surface polaritons were studied in more detail and new types were analyzed in hyperbolic media: Dyakonov plasmons~\cite{Jacob2008OpticalHyperspace,Miret2012DyakonovlikeSurface,Cojocaru2014ComparativeAnalysis,Davidovich2020DyakonovPlasmon}, DSWs~\cite{Takayama2012PracticalDyakonons}, SPP and DSW coupling on two closely spaced interfaces~\cite{Takayama2012CouplingPlasmons,Chen2014AnalysisDyakonov}, and others (see review~\cite{Takayama2017PhotonicSurface}).

\paragraph*{Experimental studies and suitable materials.}
Despite the fact that the work is devoted to a theoretical description of polaritons, it is worth mentioning experimental studies.
Surface waves have been actively studied and continue to be studied, especially in the context of near-field heat transfer~\cite{Francoeur2008NearfieldRadiative,Dai2015EnhancedNearfield,Zhang2018NearFieldRadiative,Yan2023SurfacePhonon}.
Examples of materials in which phonon polaritons of interest to us were studied: \ce{LiTaO3} and \ce{LiNbO3}~\cite{Barker1970InfraredStudy}, \ce{CdS}~\cite{Perry1973DispersionSurface}, \ce{MgF2} and \ce{TiO2}~\cite{Bryksin1973SurfaceOscillations}, \ce{BaTiO3} and \ce{PbTiO3}~\cite{Fischer1974SurfacePolaritons}, \(\alpha\)--\ce{SiO2}~\cite{Falge1973DispersionPhononLike,Falge1974DispersionGeneralized}, \( \alpha\)--\ce{LiIO3}~\cite{Puchkovskaya1978CrystalOptics}, uniaxial \ce{Al2O3}~\cite{Chu1986DispersionSurface}, 6H--\ce{SiC}~\cite{Melnichuk1998InfluenceAnisotropy}, \ce{CaCO3}~\cite{Ma2021GhostHyperbolic}; see also reviews~\cite{Borstel1977SurfacePhononpolaritons,Korzeb2015CompendiumNatural,Narimanov2015NaturallyHyperbolic}.
Polaritons in anisotropic van der Waals structures are also of great interest at the present time~\cite{Ma2020AnisotropicPolaritons}.
Surface plasmon polaritons at the interface of a metal and an anisotropic dielectric have also been studied experimentally in liquid crystal~\cite{Yang1982DeterminationLiquid}, in \ce{CaCO3}~\cite{Simon1995SurfaceElectromagnetic}, in \ce{ZnO}~\cite{Melnichuk1998SurfacePlasmonphonon}, in \textit{para}-sexiphenyl nanowire films~\cite{Takeichi2011AnisotropicPropagation}.
The possibility of creating anisotropic metal layers and, accordingly, the existence of anisotropic surface plasmons was experimentally demonstrated in~\cite{Feng2010FormBirefringence,Muskens2007ModificationPhotoluminescence}.
Hyperbolic waves and dyakonov plasmons were studied in large number of artificial structures, for example based on \ce{Au} nanorods~\cite{Kabashin2009PlasmonicNanorod}, on \ce{Ag}/\ce{SiN} thin-layered structure~\cite{Maas2013ExperimentalRealization}, on metal gratings \ce{Ag}~\cite{High2015VisiblefrequencyHyperbolic}, on \ce{Si}/\ce{ZnO}:\ce{Al} nanorods~\cite{Riley2016HighQualityUltraconformal}, deep trench structures~\cite{Takayama2017MidinfraredSurface,Takayama2018ExperimentalObservation}, nanostructured hexagonal \ce{BN}~\cite{Li2018InfraredHyperbolic}.
Surface hyperbolic phonon-polaritons in two-dimensional films are being actively studied, for example in a single \cite{Ma2018InplaneAnisotropic,Dai2020EdgeorientedSteerable} or twisted \cite{Chen2020ConfigurablePhonon} \( \alpha \)-\ce{MoO3} flakes.
At the same time, the DSWs have not been experimentally studied in such detail as others.
The first observation of DSW was made more than 20 years after the theoretical prediction in~\cite{Takayama2009ObservationDyakonov} at the interface of positive biaxial crystal (\ce{KTiOPO4}) and index--matching liquid.
There are only a few other studies in which the excitation of DSWs was also observed: nematic liquid crystal and polycarbonate~\cite{Li2020ControllableSelective}, \ce{MgF2} and chiral sculptured thin film~\cite{Pulsifer2013ObservationDyakonovTamm}.
Guided modes enabled by DSW have been observed in \( \ce{Al2O3} \) nanosheet between negative biaxial crystal (\ce{LiB3O5}) and index-matching liquid~\cite{Takayama2014LosslessDirectional}.

In all works up to \cite{Kroytor2021InvestigationExistence}, it was believed that the solution to the dispersion equation of a surface wave for an arbitrary direction of propagation cannot be obtained in analytical form.
The systems of equations were solved numerically.
This may be inconvenient for further analysis of frequency dispersion, for example in metamaterials.
Therefore, the aim of the present work is to extend the solution for Dyakonov surface waves given in \cite{Kroytor2021InvestigationExistence}, where only positive dielectric permittivities were considered, to the case of arbitrary dielectric permittivities.
A universal analytical solution in algebraic form is given for an arbitrary ratio of the dielectric permittivities of media.

The main part of the work is devoted to the detailed analysis of it and correspondence with previously known results.
This part also provides examples of the basic isofrequency contour shapes depending on the dielectric permittivities.
When analyzing combinations of different materials, it was discovered that two types of surface polaritons were not described in the literature.
The first type is polaritons at the interface of isotropic metal-like medium and Type I hyperbolic medium which can be elliptic- or hyperbolic-like.
The second is elliptic polaritons at the interface of anisotropic medium with both negative dielectric permittivities.
Other papers mentioned only hyperbolic polaritons.
The classification of surface polaritons according to the shapes of isofrequency contours is given in summary.

Some cumbersome transformations that may be useful are given in the appendices for those interested.

\section{Model}
\begin{figure}[t]
    \includegraphics{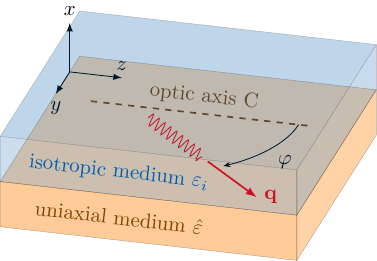}
    \caption{\label{fig:sketch}A planar interface is formed between an uniaxial medium and an isotropic medium where surface polaritons propagates at angle \( \varphi \) to the \( z \) axis. The optic axis of an uniaxial medium coincided with the \( z \) axis and parallel to the interface plane.}
\end{figure}
We are looking for surface electromagnetic waves propagating along the interface of two homogeneous non-magnetic non-chiral media.
Electromagnetic properties of media are describe within dielectric permittivity which may depend on the frequency of the wave \( \omega \).
Let us suppose that an isotropic medium filling the upper half-space \( x > 0 \) and an anisotropic uniaxial medium filling the lower half-space \( x < 0 \).
Let \( \varepsilon_i \) be dielectric permittivity of the isotropic medium and \( \hat{\varepsilon} \) be dielectric tensor of the anisotropic medium.
In principal axes
\begin{equation}\label{eq:anisotropic-medium-permittivity}
    \hat{\varepsilon} = \begin{pmatrix}
        \varepsilon_o & 0 & 0 \\
        0 & \varepsilon_o & 0 \\
        0 & 0 & \varepsilon_e
    \end{pmatrix}
\end{equation}
has two independent components \footnote{\( \varepsilon_o = \varepsilon_\perp \) and \( \varepsilon_e = \varepsilon_\parallel \) in Dyakonov's notation~\cite{Dyakonov1988NewType}}.

As a good first approximation, we neglect absorption, dissipation, and gain in media if \( \abs{\Re \varepsilon_{i,o,e}} \gg \abs{\Im{\varepsilon_{i,o,e}}} \).
Then let \( \Im \varepsilon_i = \Im \varepsilon_o = \Im \varepsilon_e = 0 \) at \( \omega \).
If we do not neglect losses then surface polariton would have finite propagation length or finite lifetime.
To find them, one can use the solution obtained as a good first approximation.
One should be careful, because such an approximation is unacceptable in \( \varepsilon \)--near zero materials in which it is assumed \( \Re \varepsilon_{i,o,e} \approx 0 \).
Also, in an anisotropic medium, this may change the AED of surface polariton.

\subsection{Dispersion equation}
Let us choose the coordinate system (Fig.~\Figref{fig:sketch}) in such a way that the plane \( x = 0 \) coincides with the interface between media.
An optic axis \( C \) is parallel to the interface and directed along \( z \) axis.
That is, the coordinate axes coincide with the principal axes of \eqref{eq:anisotropic-medium-permittivity}.
Let \( \mathbf{q} = (0, q_y, q_z) \) be wavevector of a surface polariton making angle \( \varphi \) with~\( C \).
It is then determined from the dispersion equation.
In a general case, a surface wave should be a linear combination of all eigenwaves, with some specific exception \footnote{
    Simultaneous degeneracy of eigenwaves (both homogeneous and inhomogeneous) in polarization and wavevector is possible for some directions in anisotropic media~\cite{Marchevskii1984SingularElectromagnetic,Fedorov1985SingularWaves}.
    A clear explanation for this is the defective eigen wavevectors \( \mathbf{k} \) in Maxwell equations.
    In such case we should take solution in the form \( \left[\mathbf{f}_1 + \mathbf{f}_2 (\mathbf{k} \mathbf{r})\right] \exp(i\mathbf{k}\mathbf{r} - i \omega t) \).
    This was first noted by W.~Voigt in 1902 \cite{Voigt1902BehaviourPleochroitic}.
    Now these directions are called singular axes or exceptional points, and the waves are called Voigt waves, singular waves or exceptional waves.
    }.
They all have the same \( \mathbf{q} \), but possibly different decay with distance from the interface.
As usual, eigenwaves mean solutions to Maxwell's equations with constitutive relations
\begin{align*}%\label{eq:maxwell-equations}
    &\curl{\vb{E}} = \frac{i \omega}{c} \vb{B}, &&\curl{\vb{H}} = -\frac{i \omega}{c} \vb{D}\\
    &\div{\vb{D}} = 0, &&\div{\vb{B}} = 0\\
    &D_{i}(\omega) = \hat{\varepsilon}_{ik}(\omega) E_k(\omega),  &&B_i = H_i,
\end{align*}
in the form of a monochromatic plane wave, such that \( \vb{E}, \vb{D}, \vb{H} \propto \exp(i q_y y + i q_z z - i \omega t \pm \kappa x) \).
In the isotropic medium we can freely choose the combination of two waves with orthogonal polarization, transverse electric (TE--) and transverse magnetic wave (TM--wave) for example.
There is no such freedom in the uniaxial medium, and the general solution is a combination of an ordinary and an extraordinary wave.
There is an exception to this that will be specifically mentioned.

Although, there are many works~\cite{Puchkovskaya1978CrystalOptics,Dyakonov1988NewType,Elston1990SurfacePlasmonpolaritons,Liscidini2010QuasiguidedSurface,Li2008HybridizedSurface,Alshits2020PlasmonPolaritons} that study surface waves in this case and even for an arbitrary optic axis direction~\cite{Walker1998SurfaceMode,Alshits2002DispersionlessSurface,Davidovich2020DyakonovPlasmon,Alshits2023PlasmonpolaritonInterfacea}.
Nowhere before~\cite{Kroytor2021InvestigationExistence} was an analytical solution obtained in a simple and elegant 1-parametric form.
Here we use the similar method proposed in~\cite{Kroytor2021InvestigationExistence} for DSWs to obtain analytical solution for any possible combination of signs \( \varepsilon_i, \varepsilon_o, \varepsilon_e \) \footnote{
    Moreover, it may be possible to obtain an analytical solution for an arbitrary direction of the optic axis.
    The technique for deriving the dispersion equation of SPP in~\cite{Alshits2023PlasmonpolaritonInterfacea} may be very helpful.}.

From now on, all dimensional quantities are expressed in units of the wavevector in vacuum \( k_0 = \omega / c \).
Using standard boundary conditions for the continuity of tangential components of the electric \( E_{y,z} \) and magnetic fields \( B_{y,z} \), we obtain the dispersion equation
%\begin{widetext}
    \begin{multline}\label{eq:dispersion-equation}
    \qty\Big[\kappa_i (n_z^2 - \varepsilon_o) + \kappa_o \qty(n_z^2 - \varepsilon_i)]\times\\
    \times\qty\Big[\kappa_i \varepsilon_i \qty(n_z^2 - \varepsilon_o) + \kappa_e \varepsilon_o \qty(n_z^2 - \varepsilon_i)]= \\
    = n_y^2 n_z^2 \qty(\varepsilon_i - \varepsilon_o)^2
    \end{multline}
%\end{widetext}
to which we need to add the relations for different eigenwaves
\begin{align}
    \kappa_i^2 &= n_y^2 + n_z^2 - \varepsilon_i,\label{eq:dispersion-equation-isotropic}\\
    \kappa_o^2 &= n_y^2 + n_z^2 - \varepsilon_o,\label{eq:dispersion-equation-ordinary}\\
    \kappa_e^2 &= n_y^2 + \frac{\varepsilon_e}{\varepsilon_o}n_z^2 - \varepsilon_e,\label{eq:dispersion-equation-extraordinary}
\end{align}
where \( \kappa_i \) --- TE-- and TM--wave decay constant, \( \kappa_o \) --- ordinary wave decay constant, \( \kappa_e \) --- extraordinary wave decay constant, \( n_y = q_y / k_0, n_z = q_z / k_0 \).
The system of four equations \eqref{eq:dispersion-equation}--\eqref{eq:dispersion-equation-extraordinary} contains five unknowns \( n_y, n_z, \kappa_i, \kappa_o, \kappa_e \) that need to be found.
One also need to remember that \( n_y = n_z \tan \varphi \).
The detailed derivation of dispersion equation \eqref{eq:dispersion-equation} and expressions for field components in a surface wave are given in Appendix~\ref{sec:fields-expressions}.

It should be noted that \eqref{eq:dispersion-equation} differs from the previously derived~\citep[Eq.~(8) in][]{Dyakonov1988NewType} and~\citep[Eq.~(12) in][]{Takayama2008DyakonovSurface}, but is similar to the dispersion equation for the eigenmodes in an anisotropic cylindrical waveguides~\citep[Eq.~(4) in][]{Golenitskii2020DyakonovlikeSurface}.
In Appendix~\ref{sec:angle-elimination}, dispersion equation \eqref{eq:dispersion-equation} is proven to be equivalent to the equation obtained by Dyakonov~\citep[Eq.~(9) in][]{Dyakonov1988NewType}.

\subsection{Algebraic solution}
We say that a surface polariton exists at angle \( \varphi \) if there is a solution to the system \eqref{eq:dispersion-equation}--\eqref{eq:dispersion-equation-extraordinary} that satisfies the conditions \( \kappa_i, \kappa_o, \kappa_e > 0 \) and \( n_y, n_z \) is real.
According to symmetry, the surface polariton also exists for angles \( -\varphi, \pi + \varphi, \pi - \varphi \).
Often the dispersion equation \eqref{eq:dispersion-equation-extraordinary} is solved numerically for a chosen direction making an angle \( \varphi \) with the optic axis.
But in fact, it is possible to write down the exact solution in algebraic form~\cite{Kroytor2021InvestigationExistence} using an additional variable
\begin{align}\label{eq:exact-solution}
    \kappa_o(s) &= \sqrt{-\frac{P_2(s)}{P_4(s)}}\qcomma \kappa_e(s) = s \kappa_o(s),\\
%\end{equation}
%\begin{equation}
    \kappa_i(s) &= \sqrt{\kappa_o^2(s) + \varepsilon_o - \varepsilon_i}, \label{eq:exact-solution-ki}\\
%\end{equation}
%\begin{equation}
    n_y(s) &= \sqrt{\frac{\qty(\varepsilon_e - \varepsilon_o s^2) \kappa_o^2(s)}{\varepsilon_e - \varepsilon_o}}, \label{eq:exact-solution-ny}\\
%\end{equation}
%\begin{equation}
    n_z(s) &= \sqrt{\frac{\qty(s^2 - 1) \varepsilon_o \kappa_o^2(s)}{\varepsilon_e - \varepsilon_o} + \varepsilon_o}, \label{eq:exact-solution-nz}\\
%\end{equation}
    q(s) &= \sqrt{n_y(s)^2 + n_z(s)^2}, \label{eq:exact-solution-q}\\
    \varphi(s) &= \arctan \dfrac{n_y(s)}{n_z(s)}, \label{eq:exact-solution-angle}
\end{align}
where \( P_2(s) \) and \( P_4(s) \) are polynomials of degree 2 and degree 4 defined as
\begin{align}
    &P_2(s) = (\varepsilon_i - \varepsilon_o)(\varepsilon_e + \varepsilon_o { s})^2, \label{eq:polynimial-p2}\\
    &\begin{multlined}
        P_4(s) = (\varepsilon_i + \varepsilon_o s)(1 + s) \times \\
        \times\left[\varepsilon_i - 2 \varepsilon_e + (\varepsilon_i - \varepsilon_o) s + \varepsilon_o s^2\right]. \label{eq:polynimial-p4}
    \end{multlined}
\end{align}
A detailed derivation of the solution \eqref{eq:exact-solution} is given in Appendix~\ref{sec:dispersion-solution}.
If the value \( s \) is known then all parameters of the surface polariton are determined uniquely.
In fact, the range of permissible values of \( s \) is limited and corresponds to the surface wave solution.
This range is determined only by the dielectric permittivities \( \varepsilon_i, \varepsilon_o, \varepsilon_e \).

The next section is devoted to the analysis of permissible values of \( s \) in different media, extending the results obtained in~\cite{Kroytor2021InvestigationExistence} for DSWs.

\section{Analysis of algebraic solution}
Let us consider the general case when the signs \( \varepsilon_i \), \( \varepsilon_o \), and \( \varepsilon_e \) are arbitrary.
We divide all combinations of signs into three groups: \textit{negative permittivity group}, \textit{positive permittivity group}, \textit{hyperbolic group}.
In fact, the division is quite arbitrary, because negative permittivity and hyperbolic groups have something in common with each other.
This will become clear further.

The number of cases that need to be analyzed is slightly reduced.
It follows from \eqref{eq:dispersion-equation} that there are no solutions if \( \varepsilon_i, \varepsilon_o < 0 \) regardless of \( \varepsilon_e \).

\subsection{Negative permittivity materials. SPPs, SPhPs}
The \textit{negative \( \varepsilon \) group} includes such cases where one medium has all negative components of the dielectric tensor, and other has all positive components of the dielectric tensor.
From \eqref{eq:dispersion-equation} it follows that there is no solution if all \( \varepsilon_i, \varepsilon_o, \varepsilon_e < 0 \).
Therefore there are four different cases of partnering media: anisotropic metal-like medium (\( \varepsilon_e < \varepsilon_o < 0 \) or \( \varepsilon_o < \varepsilon_e < 0 \)) and isotropic dielectric \( \varepsilon_i > 0 \); isotropic metal-like medium \( \varepsilon_i < 0 \) and uniaxial dielectric medium, positive (\( \varepsilon_o < \varepsilon_e \)) or negative (\( \varepsilon_e < \varepsilon_o \)).
It is clear that in the limit of weak anisotropy (\( \varepsilon_o \approx \varepsilon_e \)), surface polaritons in this group include surface plasmon polaritons (SPPs), surface phonon polaritons (SPhPs) and other polaritons in isotropic media with negative permittivity \footnote{
    In these cases, when \( \Im \varepsilon = 0 \), some authors~\cite{Polo2013ElectromagneticSurface} call surface waves \textit{Fano waves}, named after the one who first derived the dispersion of surface plasmon polariton~\cite{Fano1941TheoryAnomalous}.}.

\subsubsection{Isotropic metal-like medium and uniaxial dielectric\label{sub:isotropic-metal}}
Surface polaritons at the boundary of a metal or medium having negative permittivity at some frequency and an uniaxial dielectric have been considered earlier.
Surface plasmon (potential wave) dispersion equation in the quasistatic approximation (\( q \gg \omega / c \)) have been obtained in~\cite{Agranovich1965SurfaceExcitons,Hussein1969StudyElectromagnetic}.
In~\cite{Li2008HybridizedSurface,Liscidini2010QuasiguidedSurface,Liu2013LeakySurface,Moradi2018TerahertzDyakonov} authors were numerically analyzed dispersion of SPPs for an arbitrary propagation direction.
Also in~\cite{Li2008HybridizedSurface}, surface polariton AED has been obtained analytically.
The paper~\cite{Alshits2020PlasmonPolaritons} pays attention to the special case when the SPP propagation length increases significantly for some directions.
It happens due to large penetration length (\( \kappa_e \to 0 \)) in the uniaxial dielectric, physically meaning that the polariton energy is transferred mainly in the dielectric without significant losses.
The case of an arbitrary oriented optic axis have been analyzed in~\cite{Alshits2023PlasmonpolaritonInterfacea} using an iterative method.

Let
\begin{equation}\label{eq:anisotropy-strength}
    \eta = \frac{\varepsilon_e}{\varepsilon_o} - 1
\end{equation}
be the relative anisotropy strength, and
\begin{equation}\label{eq:relative-eps-i}
    \chi = -\frac{\varepsilon_i}{\varepsilon_o}
\end{equation}
be the relative permittivity of the isotropic medium.
More general \( \eta = \eta(\omega) \) and \( \chi = \chi(\omega) \).
Then \( \eta > 0 \) corresponds to positive birefringence and \( -1 < \eta < 0 \) to negative birefringence.
In this section \( \chi > 0 \).
Its value is important and allows us to make several statements about the existence of surface polaritons.
If \( \chi > 1 \) and anisotropy is weak \( \eta \approx 0 \) then a surface polariton propagating perpendicular to the optic axis \( \varphi = \pi/2 \) exists.
This is clear from the well-known dispersion law for SPP in isotropic media
\begin{equation}\label{eq:spp-dispersion}
    q_\text{SPP}^2 = \frac{\varepsilon_i \varepsilon_o}{\varepsilon_i + \varepsilon_o} = \frac{\varepsilon_o \chi}{\chi - 1}.
\end{equation}
More details is given in Appendix~\ref{sec:high-symmetry-directions}.

\begin{figure}[!]
    \includegraphics[height=0.744\textheight]{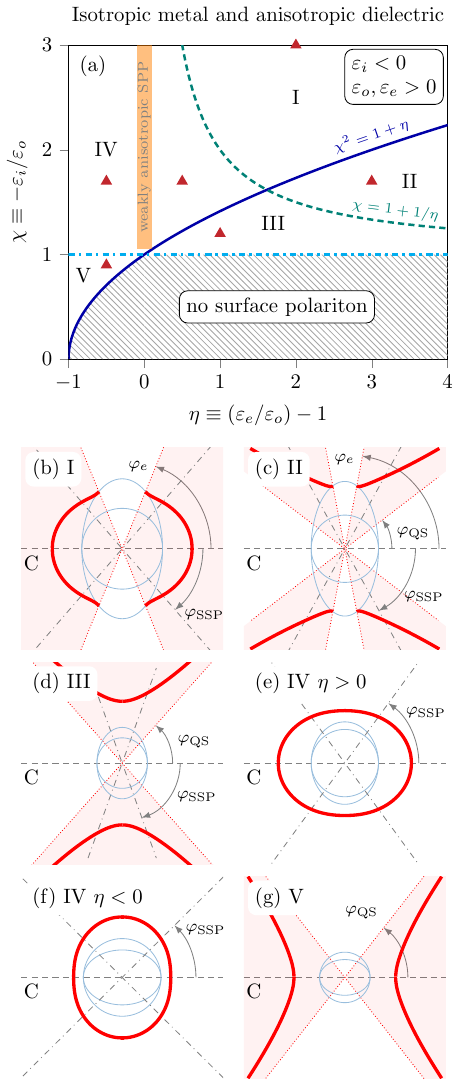}
    \caption{\label{fig:spp-types}
        Types of surface polaritons and its existence domains at the interface of a metal-like medium \( \varepsilon_i < 0 \) and an uniaxial dielectric \( \varepsilon_o, \varepsilon_e > 0 \).
        (a) Surface polariton existence domains in material parameters space.
        Vertical orange rectangle shows the domain corresponding to usual surface plasmon polariton (SPP) in a weakly anisotropic medium.
        The shaded region is the domain of parameters in which there is no surface polariton at any propagation angle \( \varphi \).
        (b)--(g) Isofrequency contours \( q(\varphi) \) of surface polariton for selected parameters (triangles) from different domains.
        The thick red curve is the isofrequency contour.
        The red shaded region is angular existence domain.
        The optic axis C is horizontal.
        Thin blue lines are the light cones for ordinary and extraordinary waves.
        Angle \( \varphi_\text{SSP} \) \eqref{eq:ssp-angle} shows the angle at which singular polariton propagates.
        Angles \( \varphi_e \) and \( \varphi_\text{QS} \) are defined by \eqref{eq:ke-zero} and \eqref{eq:q-infinite}.
    }
\end{figure}

Let us look at the exact solution \eqref{eq:exact-solution}.
The range of \( s \) is determined in this case from the condition that the radical expressions \eqref{eq:exact-solution-ny} and \eqref{eq:exact-solution-nz} are positive and finite.
When solving inequalities, it turns out to be convenient to introduce special values
\begin{align}
    s_1 &= \begin{cases}
        \sqrt{1 + \eta - \eta\chi}, &\qif* 1 + \eta - \eta\chi > 0\\
        0, &\qotherwise*
    \end{cases} \label{eq:critical-point-s1}\\
    s_2 &= \sqrt{1 + \eta}, \label{eq:critical-point-s2}\\
    s_3 &= \chi. \label{eq:critical-point-s3}
\end{align}
These values corresponds to zeros or poles of any rational expression under radicals \eqref{eq:exact-solution}--\eqref{eq:exact-solution-q}: \( n_z \to 0 \) as \( s \to s_1 > 0 \) (if \( s = 0 \) then \( \kappa_e = 0 \)); \( n_y \to 0 \) as \( s \to s_2 \); \( P_4(s) \to 0 \) as \( s \to s_3 \).
Condition \eqref{eq:critical-point-s1} may seem artificial, but it has a clear physical sense.
The extraordinary wave becomes TM-wave for \( \varphi = \pi / 2 \) and has only \( E_y, E_x, H_z \) components.
It means that surface polariton wavevector \( q \) is determined by \( \varepsilon_o \) and \( \varepsilon_i \) only \eqref{eq:spp-dispersion}.
But at the same time, for the existence of a surface wave, \( q_\text{SPP} \) must be greater than the wavevector of a homogeneous extraordinary wave \( q_\text{SPP}^2 > \varepsilon_e \), which implies \eqref{eq:critical-point-s1}.

One may note that \( s_1 < s_2 \) if \( \eta > 0 \), and \( s_1 > s_2 \) if \( \eta < 0 \).
The range of \( s \) differs slightly for these cases.
Let us consider both cases in more detail.

\paragraph*{Positive birefringent dielectric, \( \eta > 0 \).}
\begin{equation*}
    s_1 < s < \min(s_2, s_3).
\end{equation*}
Using \eqref{eq:critical-point-s1}--\eqref{eq:critical-point-s3} we obtain a necessary and sufficient condition for the existence of a surface polariton
\begin{equation*}
    \chi > 1 \quad \Leftrightarrow \quad \varepsilon_o + \varepsilon_i < 0.
\end{equation*}

One can already notice several features.
If \( s_1 = 0 \) then \( \kappa_e = 0 \) for some angle \( \varphi(0) = \varphi_e \) and
\begin{equation}\label{eq:ke-zero}
    \tan^2 \varphi_e = \frac{(1 + \eta)^3 (1 + \chi)}{(1 + \eta + \chi)(\eta \chi - 1 - \eta)}.
\end{equation}
It means that surface polariton is weakly localized in dielectric for directions \( \varphi \approx \varphi_e \) which is noted in~\cite{Li2008HybridizedSurface,Alshits2020PlasmonPolaritons}.
\( \varphi_e \) exists if \( \eta, \chi \) fall into domain \( \RN{1} + \RN{2} \) in Fig.~\Figref[(a)]{fig:spp-types}.

If \( \min(s_2, s_3) = s_3 \) and \( \kappa_o, \kappa_e, \kappa_i, q \to \infty \) as \( s \to s_3 \).
This limit corresponds to the quasistatic approximation, when \( \kappa_o, \kappa_e, \kappa_i, q \gg k_0 \) and we can neglect retardation.
The corresponding angle \( \varphi_\text{QS} \) at which this occurs is determined by
\begin{equation}\label{eq:q-infinite}
    \tan^2 \varphi_\text{QS} = \frac{1 + \eta - \chi^2}{\chi^2 - 1}.
\end{equation}
Angle \( \varphi_\text{QS} \) \footnote{\( \varphi_e = \varphi' \) and \( \varphi_\text{QS} = \varphi'' \) in notation~\cite{Li2008HybridizedSurface}} exists only if \( 1 < \chi < \sqrt{1 + \eta} \), corresponding domain \( \RN{2}+\RN{3} \) in Fig.~\Figref[(a)]{fig:spp-types}.
Analytical expression \eqref{eq:q-infinite} was first obtained in~\cite{Hussein1969StudyElectromagnetic}.
It should be understood that \( q \to \infty \) as \( \varphi \to \varphi_\text{QS} \) only in the absence of any losses (\( \Im \varepsilon_{o,e,i} = 0 \)).
Therefore, in real materials \( \Re q \) is finite, but possibly large.

\paragraph*{Negative birefringent dielectric, \( \eta < 0 \).}
\begin{equation*}
    s_2 < s < \min(s_1, s_3).
\end{equation*}
The existence condition in this case includes part of the region \( \chi < 1 \) and is written as
\begin{equation*}
    \chi > \sqrt{1 + \eta}.
\end{equation*}
If \( \chi < 1 \) then angle \( \varphi_\text{QS} \) exists in the same way as in the case of \( \eta > 0 \).
This domain is marked \RN{5} in Fig.~\Figref[(a)]{fig:spp-types}.

When \( \eta \neq 1 \) and \( \chi > 1 \) (domains \RN{1}--\RN{4}) singular surface polariton (SSP) exists which is propagating at an angle
\begin{equation}\label{eq:ssp-angle}
    \tan^2 \varphi_\text{SSP} = \frac{(2 + \eta)^2 (1 + \chi)}{4(1 + \eta + \chi) (\chi - 1)}.
\end{equation}
At this propagation angle \( \kappa_e = \kappa_o \), and the polarization vectors of the ordinary and extraordinary wave are proportional to each other.
This leads to the fact that the independent solutions of the Maxwell's equations are not \( \vb{A}_o\exp\qty({-\kappa_o x}) \) and \( \vb{A}_e\exp\qty({-\kappa_e x}) \), but \( \vb{A}_1\exp\qty({-\kappa x}) \) and \( \vb{A}_2 x \exp({-\kappa x}) \) where \( \kappa = \kappa_o = \kappa_e \) \cite{Voigt1902BehaviourPleochroitic}.
This consideration for surface waves named SSP were discussed in \cite{Marchevskii1984SingularElectromagnetic}, but only in the case of isotropic dielectric.
The case of isotropic metal and anisotropic dielectric was addressed in \cite{Zhou2019SurfaceplasmonpolaritonWave}, and the authors called this SPP--Voigt wave.

It should be noted that \( \eta = 0 \) and \( s_1 = s_2 = 1 \) for the case of isotropic media.
And accordingly, the range of \( s \) degenerates into a single point.
Having carefully calculated the limit in this case, we obtain the dispersion of an isotropic SPP \eqref{eq:spp-dispersion}.

All analysis results are shown in Fig.~\Figref{fig:spp-types}.
It shows all domains of material parameters in which surface polariton exists and examples of isofrequency contours \( q(\varphi) \) for different material parameters.
This should be understood as follows.
\( \eta \) and \( \chi \) change with frequency \( \omega \) due to frequency dispersion, and accordingly the point \( (\eta, \chi) \) in the plot (Fig.~\Figref[(a)]{fig:spp-types}) falls into different domains depending on \( \omega \).
This leads to a change in the shape of the polariton isofrequency contour and AED.
The combined existence condition can be written as
\begin{equation*}
    \chi > \min(1, \sqrt{1 + \eta}).
\end{equation*}
It is equivalent to the existence condition given in~\cite{Li2008HybridizedSurface}.
The case of isotropic media falls into domain \RN{4}.
In this domain of parameters the contour is ellipse-like and is most similar to a circle as for an isotropic SPP.
AED in \RN{4} is a full plane, \( \varphi \in [0, 2\pi) \).
In domain \RN{1} the angle \( \varphi_e \) exists and the surface polariton contour looks like the trimmed ellipse.
If material parameters falls into domains \RN{2}, \RN{3}, and \RN{5} then the contour is hyperbola-like curve due to the existence of angle \( \varphi_\text{QS} \).
Depending on the type of isofrequency curve, we call surface polaritons \textit{elliptic-like} or \textit{hyperbolic-like}.
It should be noted that only in domain \RN{2} the surface polariton do not exist for highly symmetric directions, along the optic axis and/or perpendicular to the optic axis.

\subsubsection{Isotropic dielectric and anisotropic metal-like medium}
\begin{figure}[t]
    \includegraphics[]{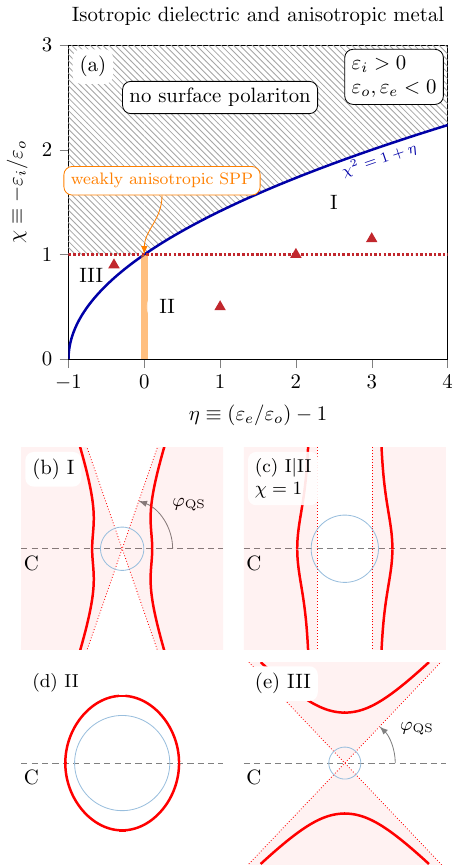}
    \caption{\label{fig:spp-types-anisotropic-metal}
        Types of surface polaritons and its existence domains at the interface of an isotropic dielectric \( \varepsilon_i > 0 \) and an uniaxial metal-like medium \( \varepsilon_o, \varepsilon_e < 0 \).
        %        (a) The same notations are used as in Fig.~\Figref{fig:spp-types}.
        (a) Surface polariton existence domains in material parameters space.
        %        Vertical orange rectangle shows the domain corresponding to usual surface plasmon polariton (SPP) in a weakly anisotropic medium.
        %        The shaded region is the domain of parameters in which there is no surface polariton at any propagation angle \( \varphi \).
        (b)--(e) Isofrequency contours \( q(\varphi) \) of surface polariton for selected parameters (triangles) from different domains.
        %        The thick red curve is the isofrequency contour.
        %        The red shaded region is angular existence domain.
        %        The optic axis C is horizontal.
        Thin blue line is the light cone in isotropic medium.
        Other notations used are the same as in Fig.~\Figref{fig:spp-types}.
        Angle \( \varphi_\text{QS} \) is defined by \eqref{eq:q-infinite}.
    }
\end{figure}

There is not much variety of surface polaritons in this case.
In the first works~\cite{Agranovich1965SurfaceExcitons,Lyubimov1972SurfaceElectromagnetic,Agranovich1975CrystalOptics} existence conditions of a surface polariton were analyzed in the quasistatic approximation and without it for directions of high symmetry.
The case of optic axis perpendicular to the boundary is analyzed in \cite{Warmbier2012SurfacePlasmon}.
SPP in anisotropic metals with anisotropic plasma frequency were studied in~\cite{Ishida1987SurfacePlasmon}.
A detailed analysis of different types of surface polaritons in this case, and its AED, was carried out in~\cite{Repan2020HyperbolicSurface} and in unpublished paper~\cite{Takayama2015HyperbolicPlasmons}.
It is also worth noting the work \cite{Liu2013MetasurfacesManipulating} in which similar results were obtained for a metal metasurface designed as diffraction grating.

It is possible to use previously defined \( \eta \) and \( \chi \) in \eqref{eq:anisotropy-strength} and \eqref{eq:relative-eps-i} to describe domain of material parameters where surface polariton exist.
The surface polariton existence condition in this case~\cite{Repan2020HyperbolicSurface} is
\begin{equation*}
    0 < \chi < \max (1, \sqrt{1 + \eta}).
\end{equation*}
The range of the parameter \( s \) again depends on the sign of the anisotropy strength \( \eta \).
\paragraph*{Positive anisotropy, \( \eta > 0 \):}
\begin{equation}
    \max(s_1, s_3) \leqslant s \leqslant s_2.
\end{equation}
\paragraph*{Negative anisotropy, \( \eta < 0 \):}
\begin{equation}
    \max(s_2, s_3) \leqslant s \leqslant s_1.
\end{equation}
Limit values \( s_{1,2,3} \) are the same \eqref{eq:critical-point-s1}--\eqref{eq:critical-point-s3}.
Similar to the previous case \( s_1 = s_2 \) in the limit of isotropic metal \( \eta = 0 \).
Carefully calculating polariton wavevector in this limit, we obtain an isotropic SPP \eqref{eq:spp-dispersion}.

All possible types of surface polaritons and material parameters domains are shown in Fig.~\Figref{fig:spp-types-anisotropic-metal}.
Surface polariton has \textit{hyperbolic-like} isofrequency contour for parameters from domains \RN{1} and \RN{3}.
The boundaries of AED are determined by the same angle \( \varphi_\text{QS} \) \eqref{eq:q-infinite}.
It can also be noted that isofrequency contour have additional inflection points in domain \RN{1}.
In domain \RN{2} surface polariton is \textit{elliptic-like}.

\subsection{Positive permittivity materials. DSWs}
The next group of materials is \textit{positive permittivity group}.
Surface polariton, known as DSW, at the interface of two media with positive components of dielectric exists only if \( 0 < \varepsilon_o < \varepsilon_i < \varepsilon_e \) \cite{Dyakonov1988NewType}.
That is, an uniaxial medium have to be positive with an additional condition to isotropic dielectric permittivity \( \varepsilon_i \).
In our notation DSW existence condition is written as
\begin{equation*}
    -1 - \eta < \chi < -1.
\end{equation*}
This can be seen as an extension of the case (Sec.~\ref{sub:isotropic-metal}) to the region \( \chi < 0 \).
Complete analytical solution to this case are discussed in~\cite{Kroytor2021InvestigationExistence}.
Let us repeat it briefly for completeness.
The parameter \( s \) varies within
\begin{equation*}
    0 < s < s_4,
\end{equation*}
where \( s_4 \) is determined from \( \kappa_i = 0 \) and equals
\begin{equation}\label{eq:critical-point-s4}
    s_4 = \frac{\chi + \sqrt{(2 + \chi)^2 + 4 \eta}}{2}.
\end{equation}
Let \( \varphi_i \) be the angle at which \( \kappa_i = 0 \).
AED is determined by the angles \( \varphi_i < \varphi < \varphi_e \), and the mirrored domains relative to optic axis and normal to it.
Angle \( \varphi_e \) is defined above \eqref{eq:ke-zero}, and \( \varphi_i \) in our notation
\begin{equation}\label{eq:ki-zero}
    \sin^2 \varphi_i = \frac{(-1-\chi)\qty(2 + \chi + \sqrt{(2 + \chi)^2 + 4\eta})}{2\eta}.
\end{equation}

\begin{figure}[t]
    \includegraphics[width=\columnwidth]{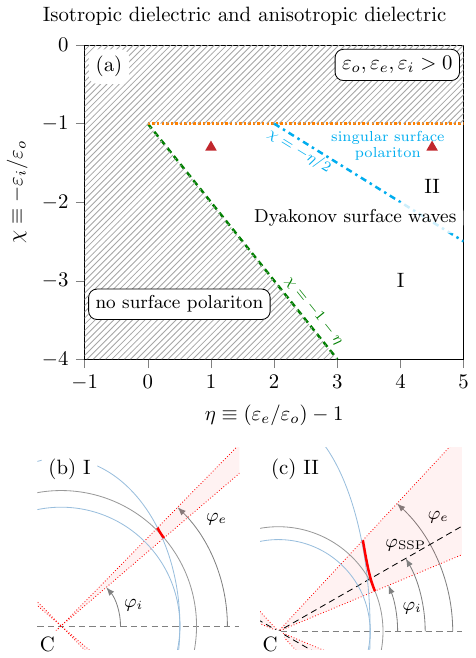}
    \caption{\label{fig:dsw-polariton}
        Dyakonov surface waves (DSWs) and singular surface polariton existence domains at the interface of an isotropic \( \varepsilon_i > 0\) and an uniaxial dielectric \( \varepsilon_o, \varepsilon_e > 0 \).
        (a) Its existence domains in material parameters space.
        The shaded region is the domain of parameters in which there is no surface polariton at any propagation angle \( \varphi \).
        (b)--(c) Isofrequency contours \( q(\varphi) \) of DSW for selected parameters (triangles) for a ``weakly'' and strongly anisotropic medium.
        %        The same notations are used as in Fig.~\Figref{fig:spp-types}.
        %        The thick red curve is the isofrequency contour.
        %        The red shaded region is angular existence domain.
        %        The optic axis (C) is horizontal.
        Thin blue lines are the light cones for ordinary, extraordinary, TE-, and TM- waves.
        Angle \( \varphi_\text{SSP} \) \eqref{eq:ssp-angle} shows the angle at which singular polariton propagates.
        Angles \( \varphi_e \) and \( \varphi_i \) are defined by \eqref{eq:ke-zero} and \eqref{eq:ki-zero}.
        Other notations used are the same as in Fig.~\Figref{fig:spp-types}.
    }
\end{figure}

Let us recall that before DSWs, the possibility of SSP existence was mentioned in \cite{Marchevskii1984SingularElectromagnetic}.
It is now often called the Dyakonov--Voigt surface wave (DVSW).
The domain of materials parameters for which it exists lies within the domain of DSWs \(  -\eta/2 < \chi < -1 \).
This was also noted in \cite{Zhou2021SingularExistence}.
Figure~\Figref{fig:dsw-polariton} shows the existence domain of DSWs (\RN{1} and \RN{2}) and DVSW (\RN{2}).

Unlike surface polaritons discussed above, DSWs are weakly localized at the interface at any propagation angle.
This can be understood by the proximity of the isofrequency contour (Fig.~\Figref[(b)--(c)]{fig:dsw-polariton}) to the light cones.
The field distributions fall away from the interface \( {\propto\,\exp(-\abs{x} / x_\text{DSW})} \), where \( x_\text{DSW} \) is the characteristic localization length.
The estimated value \( x_\text{DSW} \gtrapprox 2\sqrt{2} \eta^{-3/2} \lambda_i \) in the limit of weak anisotropy \( \eta \ll 1 \), where \( \lambda_i \) is wavelength in isotropic medium \( \lambda_i \).
For example, \( \eta \approx 0.8 \) in strongly anisotropic \ce{Hg2Cl2} for visible wavelengths and the corresponding length \( x_\text{DSW} \approx 10 \lambda_i \).
In more common natural anisotropic minerals \( x_\text{DSW} \approx {10-100 \lambda_i} \).

\subsection{Hyperbolic medium}
In the last \textit{hyperbolic group} we include cases in which the uniaxial medium has different signs of the dielectric tensor components.
Indefinite medium or hyperbolic medium \cite{Poddubny2013HyperbolicMetamaterials,Shekhar2014HyperbolicMetamaterials}, as we now call them, have actually been known for a long time in the infrared.
The contribution of optical phonons to the dielectric permittivity of a uniaxial crystal can be highly anisotropic, leading to hyperbolic dispersion of electromagnetic waves in the bulk.
Hyperbolic dispersion means that the light cone of an extraordinary wave is a one-sheet or two-sheet hyperboloid, instead of an ellipsoid.
And therefore, researchers were also interested in the dispersion of surface polaritons in such a hyperbolic medium.
The analysis of surface polariton was first done in the quasi-static approximation in \cite{Hussein1969StudyElectromagnetic,Lyubimov1972SurfaceElectromagnetic}, and then revisited much later \cite{Moradi2020ElectrostaticDyakonovlike} in structured hyperbolic metamaterials.
For a long time, the exact solution was known only for directions of high symmetry \cite{Agranovich1975CrystalOptics}, and in the form of a singular surface polariton \cite{Marchevskii1984SingularElectromagnetic}.
Researchers' interest in surface polaritons in already known hyperbolic media returned after a long time.
In \cite{Jacob2008OpticalHyperspace} it was proposed to exploit hyperbolic metamaterials for extending AED of DSWs.
The resulting surface polariton is called Dyakonov plasmon because it ``combines'' the properties of both SPP and DSW.

It is known that there are two types of hyperbolic media \cite{Shekhar2014HyperbolicMetamaterials}.
In Type \RN{1} hyperbolic medium two components of dielectric tensor are positive \( \varepsilon_o > 0 \), and one is negative \( \varepsilon_e < 0 \).
In Type \RN{2} hyperbolic medium two components are negative \( \varepsilon_o < 0 \), one is positive \( \varepsilon_e > 0 \).
Let us consider this cases separately.

\subsubsection{Type I hyperbolic medium}
\begin{figure}[!]
    \includegraphics[height=0.79\textheight]{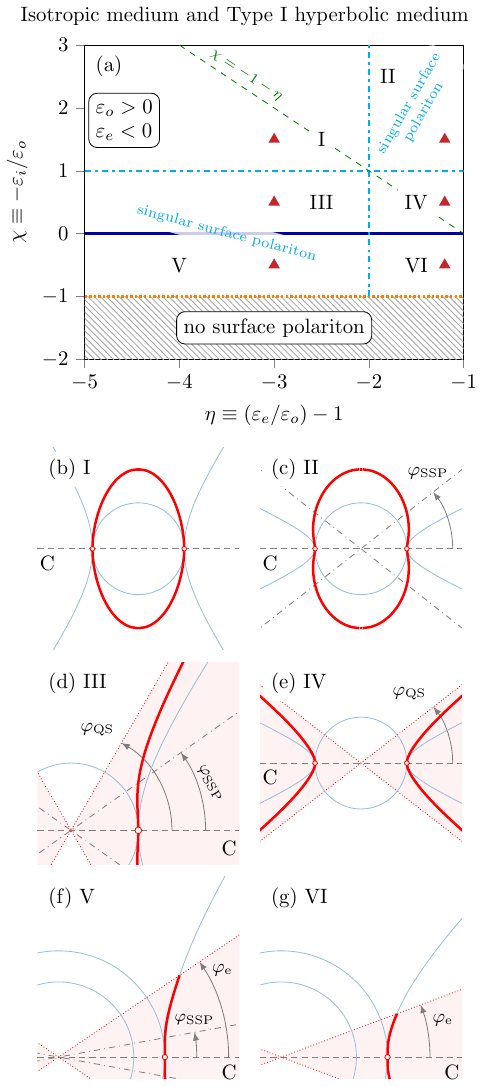}
    \caption{\label{fig:hyperbolic-1-polaritons}
        Types of surface polaritons and its existence domains at the interface of an isotropic medium \( \varepsilon_i \) and a Type \RN{1} hyperbolic medium \( \varepsilon_o > 0 \), \( \varepsilon_e < 0 \).
        (a) Surface polariton and singular surface polariton existence domains in material parameters space.
        %        The shaded region is the domain of parameters in which there is no surface polariton at any propagation angle \( \varphi \).
        (b)--(g) Isofrequency contours \( q(\varphi) \) of surface polariton for selected parameters (triangles) from different domains.
        %        The thick red curve is the isofrequency contour.
        White point at the contour shows that surface polariton can not propagate along the optic axis.
        %        The red shaded region is angular existence domain.
        %        The optic axis (C) is horizontal.
        Thin blue lines are the light cones for ordinary, extraordinary, TE-, and TM- waves.
        Angle \( \varphi_\text{SSP} \) \eqref{eq:ssp-angle} shows the angle at which singular polariton propagates.
        Angles \( \varphi_e \) and \( \varphi_\text{QS} \) are defined by \eqref{eq:ke-zero} and \eqref{eq:q-infinite}.
        Other notations used are the same as in Fig.~\Figref{fig:spp-types}.
    }
\end{figure}

To the best of author's knowledge, this case is the least well described in the literature and may not have been studied experimentally.
Since \( \varepsilon_o > 0 \), the existence of a surface polariton at the boundary is possible not only with an isotropic dielectric (e.g. air), but also with a metal-like medium.
In our notation \( \chi < 0 \) corresponds to the case of a dielectric and \( \chi < 0 \) to a ``metal''.

The latter case means, in a sense, the coupling of SPP in a metal and a surface phonon polariton (SPhP) in a hyperbolic medium.
It should be distinguished from the case of SPP--SPhP coupling in the same medium \cite{Wendler1986LongrangeSurface,Melnichuk1998SurfacePlasmonphonon}.
The first description of the surface polariton at the interface of a metal and a hyperbolic medium is given in the quasi-static approximation \cite{Gurevich1975SurfacePlasmonpolaritons,Tarkhanian1975TheorySurface}.
Much later and independently, the possibility of the existence of the surface waves in a planar anisotropic waveguide with perfectly conducting walls was noted in \cite{Podolskiy2005StronglyAnisotropic}.
It is possible that hybrid SPPs in graphene coupled anisotropic vad~der~Waals material \cite{Hajian2021HybridSurface} refers specifically to the case under discussion in the limit of thin metal layer.

The case of a boundary with a dielectric was first considered in \cite{Marchevskii1984SingularElectromagnetic} (free surface \ce{LiTaO3}), in the form of a singular surface polariton propagating at fixed angle \( \varphi_\text{SSP} \).
A more detailed analysis for an arbitrary propagation angles was carried out in \cite{Zapata-Rodriguez2013EngineeredSurface,Takayama2012PracticalDyakonons,Cojocaru2014ComparativeAnalysis} by numerically solving dispersion equation.

The only necessary and sufficient condition for surface polariton existence in this case is \( \varepsilon_i < \varepsilon_o \) or in our notation
\begin{equation}
    \chi > -1.
\end{equation}
The range of acceptable values \( s \) in the exact solution \eqref{eq:exact-solution} in the case under consideration is not as simple as the previous ones.
To describe the range of possible values of \( s \) in this case, another special value is needed
\begin{equation}\label{eq:critical-point-s5}
    s_5 = -1 - \eta.
\end{equation}

First of all, it should be noted that there is an exact solution that has a very simple form.
If it turns out that \( \varepsilon_i = \varepsilon_e \), and therefore \( \chi = - 1 - \eta \), then dispersion equation \eqref{eq:dispersion-equation} can be simplified to the form
\begin{equation}\label{eq:simplified-dispersion-equation}
    \varepsilon_i \kappa_o + \varepsilon_o \kappa_e = 0.
\end{equation}
The solution to it is
\begin{equation}\label{eq:strange-exact-solution}
    n_y^2\,\frac{\eta + 2}{\eta + 1} + n_z^2 = \varepsilon_o.
\end{equation}
The shape of isofrequency contour is an ellipse if \( \eta < -2 \), and a hyperbola otherwise.
In terms of \( s \) the range of possible values degenerates to a point \( s = s_1 = s_3 = s_4 = s_5 \).
This is similar to the isotropic case \( \varepsilon_o = \varepsilon_e \).
It can also be noted that \( P_2(s) \) and \( P_4(s) \) have a common divisor when \( \varepsilon_i = \varepsilon_e \).

In all other cases, the range of \( s \) is not degenerated.
For \( \chi > -1 - \eta \) it is defined by
\begin{equation*}
    s_5 < s < \min(s_1, s_3),
\end{equation*}
and in the region \( \chi < -1 - \eta \) by
\begin{equation*}
    \max(s_3, s_1) < s < s_5.
\end{equation*}

There are also many types of isofrequency contours of surface polariton (Fig.~\Figref{fig:hyperbolic-1-polaritons}), as in the case of a metal and an anisotropic dielectric (Sec~\ref{sub:isotropic-metal}).
They all have one feature in common.
The propagation of polariton along the optic axis is prohibited, but at an any small angle \( \varphi \neq 0 \) it is already possible.

As far as the author knows, only the case of the boundary between a dielectric and a hyperbolic medium has been considered before.
In the notation used, this means \( \chi < 0 \).
These are domains \RN{5} and \RN{6} in Fig.~\Figref[(a)]{fig:hyperbolic-1-polaritons}.
This case corresponds, for example, to a surface polariton at the free boundary (\( \varepsilon_i = 1 \)) of hyperbolic medium.
AED of a surface polariton in this case is limited by angle \( \varphi_e \) \eqref{eq:ke-zero} \cite{Zapata-Rodriguez2013EngineeredSurface,Cojocaru2014ComparativeAnalysis}.
Before these works, the existence of a SSP in domain \RN{5} was predicted \cite{Marchevskii1984SingularElectromagnetic}.
It propagates at angle \( \varphi_\text{SSP} \) \eqref{eq:ssp-angle}.
The isofrequency contour shape is more like an arc than an ellipse or a hyperbola.
This is reminiscent of DSWs.
The polariton in this case is also weakly localized.
This can be understood by the proximity of the isofrequency contour (Fig.~\Figref[(f)--(g)]{fig:hyperbolic-1-polaritons}) to the light cone of extraordinary waves.

\begin{figure}[b]
    \includegraphics[width=\columnwidth]{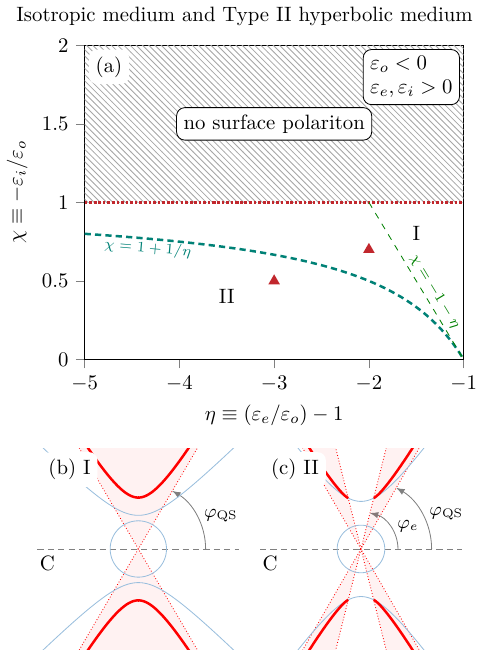}
    \caption{
        Types of surface polaritons and its existence domains at the interface of an isotropic dielectric \( \varepsilon_i > 0 \) and a Type \RN{2} hyperbolic medium \( \varepsilon_o < 0 \), \( \varepsilon_e > 0 \).
        (a) Surface polariton existence domain in material parameters space.
        %        The shaded region is the domain of parameters in which there is no surface polariton at any propagation angle \( \varphi \).
        (b)--(c) Isofrequency contours \( q(\varphi) \) of surface polariton for selected parameters (triangles) from different domains.
        %        The thick red curve is the isofrequency contour.
        %        The red shaded region is angular existence domain.
        %        The optic axis (C) is horizontal.
        Thin blue lines are the light cones for extraordinary, TE-, and TM- waves.
        Angles \( \varphi_e \) and \( \varphi_\text{QS} \) are defined by \eqref{eq:ke-zero} and \eqref{eq:q-infinite}.
        Other notations used are the same as in Fig.~\Figref{fig:spp-types}.
    }
    \label{fig:hyperbolic-2-polaritons}
\end{figure}

\begin{table*}
    \caption{\label{tab:polariton-classification}List of surface polaritons isofrequency contour shapes and its basic properties.}
    \renewcommand{\arraystretch}{1.75}
    \begin{NiceTabular}{@{\hspace{0.2cm}}*{3}{wc{0.4cm}}*{6}{@{\hskip 0.4cm}l}@{\hspace{0.2cm}}}[
        tabularnote={\raggedright \( \eta = \varepsilon_e / \varepsilon_o - 1 \); \( s_1, s_2, s_3, s_4, s_5 \) are defined by Eqs.~\eqref{eq:critical-point-s1}--\eqref{eq:critical-point-s3}, \eqref{eq:critical-point-s4}, \eqref{eq:critical-point-s5}},
        cell-space-limits=2pt
        ]
        \CodeBefore
        \rowcolor{gray!10}{3,9,10,11,13,14,16,17}
        \Body
        \toprule
        \Block{1-3}{Sign} & & & \Block{2-1}{Contour shape} & \Block{2-1}{\( s \) range\\in Eqs.\eqref{eq:exact-solution}--\eqref{eq:exact-solution-angle}} & \Block{2-1}{AED\tabularnote{angular existence domain (quadrant I only)}} & \Block{2-1}{Existence condition} & \Block{2-1}{Refs.\tabularnote{In all references below the dispersion equation~\eqref{eq:dispersion-equation} of surface polariton is solved numerically. An analytical solution in particular cases is given in \cite{Dyakonov1988NewType,Alshits2020PlasmonPolaritons,Kroytor2021InvestigationExistence,Alshits2023PlasmonpolaritonInterfacea}.}} & \Block{2-1}{SSP\tabularnote{singular surface polariton}} \\
        \cmidrule(l{0.2cm}r{0.5cm}){1-3}
        \( \varepsilon_i \)  &  \( \varepsilon_o \) &  \( \varepsilon_e \) & & & & & \\
        \midrule
        \( - \)  &  \( - \)  &  \( \pm \)  &  --- & --- & --- & --- & --- &  --- \\
        \( - \) &  \( + \)  &  \( + \)  & ``ellipse'' & \( [s_1, s_2] \cup [s_2, s_1]\tabularnote{\( s = s_1 = s_2 \) corresponds to the isotropic case \( \varepsilon_o = \varepsilon_e \)} \) & \( \left[0, \dfrac{\pi}{2}\right] \) & \Block[l,t]{}{\( \varepsilon_e < \varepsilon_o < \abs{\varepsilon_i} \) or \\ \( \sqrt{\varepsilon_o\varepsilon_e} < \abs{\varepsilon_i} < \dfrac{\varepsilon_e}{\eta} \); \tabularnote{Last inequality can be read as \( \varepsilon_e < \varepsilon_o \abs{\varepsilon_i} \qty(\abs{\varepsilon_i} - \varepsilon_o)^{-1} \), simply meaning that the polariton wavevector greater than the greatest possible wavevector for an extraordinary wave.}} & \cite{Li2008HybridizedSurface,Liscidini2010QuasiguidedSurface,Alshits2020PlasmonPolaritons,Alshits2023PlasmonpolaritonInterfacea} & yes \cite{Zhou2019SurfaceplasmonpolaritonWave} \\[0.55cm]
        & & & ``elliptic'' arc & \( (0, s_2] \) & \( \left[0, \varphi_e\right) \) & \( \max\qty(\dfrac{\varepsilon_e}{\eta}, \sqrt{\varepsilon_o\varepsilon_e}) < \abs{\varepsilon_i} \) & \cite{Li2008HybridizedSurface,Alshits2020PlasmonPolaritons,Liscidini2010QuasiguidedSurface,Alshits2023PlasmonpolaritonInterfacea} & yes \cite{Zhou2019SurfaceplasmonpolaritonWave} \\
        & & & ``\( 0 \)--hyperbola'' & \( [s_2, s_3) \) & \( \left[0, \varphi_\text{QS}\right) \) & \( \sqrt{\varepsilon_o\varepsilon_e} < \abs{\varepsilon_i} < \varepsilon_o \) & \cite{Li2008HybridizedSurface} & no \cite{Zhou2019SurfaceplasmonpolaritonWave} \\
        & & & ``\( \pi/2 \)--hyperbola'' & \( [s_1, s_3) \) & \( \left(\varphi_\text{QS}, \dfrac{\pi}{2}\right] \) & \( \varepsilon_o < \abs{\varepsilon_i} < \min\qty(\dfrac{\varepsilon_e}{\eta}, \sqrt{\varepsilon_o\varepsilon_e}) \) & \cite{Li2008HybridizedSurface} & yes \cite{Zhou2019SurfaceplasmonpolaritonWave} \\
        & & & \Block[l,t]{}{``\( \pi/2 \)--hyperbolic''\\infinite arc} & \( (0, s_3) \) & \( (\varphi_\text{QS}, \varphi_e) \) & \( \dfrac{\varepsilon_e}{\eta} < \abs{\varepsilon_i} < \sqrt{\varepsilon_o \varepsilon_e} \) & \cite{Li2008HybridizedSurface} & yes \cite{Zhou2019SurfaceplasmonpolaritonWave} \\[0.4cm]
        \( + \) & \( - \) & \( - \) & ``ellipse'' & \( [s_1, s_2] \cup [s_2, s_1]\tabularnote{\( s = s_1 = s_2 \) corresponds to the isotropic case \( \varepsilon_o = \varepsilon_e \)} \) & \( \left[0, \dfrac{\pi}{2}\right] \) & \( \varepsilon_i < \min\qty(\abs{\varepsilon_o}, \sqrt{\abs{\varepsilon_o} \abs{\varepsilon_e}}) \) & this work & no \\
        & & & ``\( 0 \)--hyperbola'' & \( (s_3, s_2] \) & \( [0, \varphi_\text{QS}) \) & \( \abs{\varepsilon_o} < \varepsilon_i < \sqrt{\abs{\varepsilon_o} \abs{\varepsilon_e}} \) & \cite{Repan2020HyperbolicSurface} & no \\
        & & & ``\( \pi/2 \)--hyperbola'' & \( (s_3, s_1] \) & \( \left(\varphi_\text{QS}, \dfrac{\pi}{2}\right] \) & \( \sqrt{\abs{\varepsilon_o} \abs{\varepsilon_e}}< \varepsilon_i < \abs{\varepsilon_o} \) & \cite{Repan2020HyperbolicSurface} & no \\
        \( + \)  &  \( + \)  &  \( + \)  & arc & \( (0, s_4) \) & \( (\varphi_i, \varphi_e) \) & \( \varepsilon_o < \varepsilon_i < \varepsilon_e \) & \cite{Dyakonov1988NewType,Kroytor2021InvestigationExistence} & \Block[l,t]{}{yes \cite{Marchevskii1984SingularElectromagnetic},\\if \( \varepsilon_e > 3 \varepsilon_o \)} \\[0.4cm]
        \( - \) & \( + \) & \( - \) & ``ellipse'' & \( [s_1, s_5) \cup (s_5, s_1]\tabularnote{\( s = s_1 = s_5 = s_3 \) corresponds to the case \( \varepsilon_e = \varepsilon_i \) \eqref{eq:strange-exact-solution}} \) &  \( \left(0, \dfrac{\pi}{2}\right] \) & \( 0 < \varepsilon_o < \abs{\varepsilon_i} \) & this work & \Block[l,t]{}{yes,\\if \( \varepsilon_o > \abs{\varepsilon_e} \)} \\[0.4cm]
        & & & ``\( 0 \)--hyperbola'' & \( (s_3, s_5) \cup (s_5, s_3)\tabularnote{\( s = s_1 = s_5 = s_3 \) corresponds to the case \( \varepsilon_e = \varepsilon_i \) \eqref{eq:strange-exact-solution}} \) & \( (0, \varphi_\text{QS}) \) & \( \abs{\varepsilon_i} < \varepsilon_o \) & this work & \Block[l,t]{}{yes,\\if \( \varepsilon_o < \abs{\varepsilon_e} \)} \\[0.4cm]
        \( + \) & \( + \) & \( - \) & \Block[l,t]{}{``\( 0 \)--hyperbolic''\\finite arc} & \( (0, s_5) \) &  \( (0, \varphi_e) \) & \( 0 < \varepsilon_i < \varepsilon_o \) & \cite{Zapata-Rodriguez2013EngineeredSurface,Cojocaru2014ComparativeAnalysis} & \Block[l,t]{}{yes \cite{Marchevskii1984SingularElectromagnetic},\\if \( \varepsilon_o < \abs{\varepsilon_e} \)}\\[0.4cm]
        \( + \) & \( - \) & \( + \) & ``\( \pi/2 \)--hyperbola'' & \( [s_1, s_3) \cup (s_3, s_1]\tabularnote{\( s = s_1 = s_5 = s_3 \) corresponds to the case \( \varepsilon_e = \varepsilon_i \) \eqref{eq:strange-exact-solution}} \) & \( \left(\varphi_\text{QS}, \dfrac{\pi}{2}\right] \) & \( \dfrac{\varepsilon_e}{\abs{\eta}} < \varepsilon_i < \abs{\varepsilon_o} \) & \cite{Jacob2008OpticalHyperspace,Takayama2012PracticalDyakonons,Cojocaru2014ComparativeAnalysis} & no \\
        & & & \Block[l,t]{}{``\( \pi/2 \)--hyperbolic''\\infinite arc} & \( (0, s_3) \) & \( (\varphi_\text{QS}, \varphi_e) \) & \( 0 < \varepsilon_i < \dfrac{\varepsilon_e}{\abs{\eta}} \) & \cite{Jacob2008OpticalHyperspace,Takayama2012PracticalDyakonons,Cojocaru2014ComparativeAnalysis} & no \\[0.3cm]
        \botrule
    \end{NiceTabular}
\end{table*}

The region \( \chi > 0 \) (domains \RN{1}--\RN{4}) correspond to the boundary with an isotropic metal-like medium and are perhaps analyzed for the first time in this paper without the quasi-static approximation.
In domains \RN{1} and \RN{2} isofrequency contours (Fig.~\Figref{fig:hyperbolic-1-polaritons}) of the surface polariton is similar to an ellipse, and we call it \textit{elliptic-like}.
AED is the entire plane except \( \varphi = 0 \) and \( \varphi = \pi \).
In domains \RN{3} and \RN{4} the surface polariton is \textit{hyperbolic-like}, as can be seen from the shape of the isofrequency contour (Fig.~\Figref[(d)--(e)]{fig:hyperbolic-1-polaritons}).
AED in this case is limited by the angle \( \varphi_\text{QS} \) \eqref{eq:q-infinite}.
The angle \( \varphi_\text{QS} \) was obtained in the quasi-static approximation in \cite{Gurevich1975SurfacePlasmonpolaritons}.
Compared to the previous case, the polariton is well localized at the boundary for any propagation direction not near the optic axis.
Also in domains \RN{2} and \RN{3} there is SSP at angle \( \varphi_\text{SSP} \) \eqref{eq:ssp-angle}.
It is worth recalling that there is a simple shape (ellipse or hyperbola) of the isofrequency contour \eqref{eq:strange-exact-solution} in \RN{1} and \RN{4} if \( \varepsilon_i = \varepsilon_e \), or \( \chi = -1 - \eta \) in our terms.

\subsubsection{Type II hyperbolic medium}
In Type \RN{2} hyperbolic medium \( \varepsilon_o < 0 \) and \( \varepsilon_e > 0 \).
As mentioned at the beginning, there is no surface polariton if \( \varepsilon_i < 0 \) and \( \varepsilon_o < 0 \).
Therefore, we only need to consider the case \( \varepsilon_i > 0 \) corresponding to \( \chi > 0 \).
The properties of surface polaritons in this case are known and well described in the literature \cite{Jacob2008OpticalHyperspace,Takayama2012PracticalDyakonons,Cojocaru2014ComparativeAnalysis,Davidovich2020DyakonovPlasmon}.

The necessary condition for existence in this case is quite simple \( \varepsilon_i < \abs{\varepsilon_o} \) which is
\begin{equation*}
    0 < \chi < 1
\end{equation*}
in our notation.
The range of possible values \( s \) is defined by
\begin{equation}
    \min\qty(s_1, s_3) < s < \max\qty(s_1, s_3),
\end{equation}
and it always includes bound \( s_1 \neq 0 \).
If \( s_1 = s_3 \), the range degenerates to a point and solution is given by Eq.~\eqref{eq:strange-exact-solution}, the same as in the case of type I hyperbolic medium.
The corresponding relation \( \chi = -1 - \eta \) is shown in Fig.~\Figref[(a)]{fig:hyperbolic-2-polaritons}.
Since \( -2 < \eta < -1 \), the shape of isofrequency contour is a hyperbola.

There are two domains of parameters in which the isofrequency contour has a different shape (Fig.~\Figref[(a)]{fig:hyperbolic-2-polaritons}).
They are separated by the relation \( \varepsilon_i = \abs{\varepsilon_o} \varepsilon_e / \qty(\abs{\varepsilon_o} + \varepsilon_e) \) which is written as \( \chi = 1 + 1 / \eta \) in the notation used.
AED of surface polariton in domain~\RN{1} (Fig.~\Figref[(b)]{fig:hyperbolic-2-polaritons}) is limited only by the angle \( \varphi_\text{QS} \) \eqref{eq:q-infinite}.
But in domain~\RN{2} there is no surface polariton propagating perpendicular to the optic axis \( \phi = \pi / 2 \), and AED is more narrow \( \varphi_\text{QS} < \varphi < \varphi_e \) (Fig.~\Figref[(c)]{fig:hyperbolic-2-polaritons}).
Isofrequency contour of polariton stars from the light cone of extraordinary waves at the angle \( \varphi_e \) \eqref{eq:ke-zero}.
In both cases surface polariton is called \textit{hyperbolic-like}.

\begin{table*}
    \caption{\label{tab:experiments}List of experimental works, media and wavelengths by type of surface polaritons.}
    \renewcommand{\arraystretch}{2}
    \begin{NiceTabular}
        {@{\hspace{0.2cm}}*{3}{wc{0.4cm}}@{\hspace{0.5cm}}l@{\hspace{1cm}}l@{\hspace{0.5cm}}l@{\hspace{0.5cm}}r@{\hspace{0.5cm}}l@{\hspace{0.2cm}}}[
            cell-space-limits=2pt
        ]
        \CodeBefore
        \rowcolor{gray!10}{6,7,8,11,12,14,15,16,17,18}
        \Body
        \toprule
        \Block{1-3}{Sign} & & & \Block{2-1}{\( (\eta, \chi) \) domain} & \Block{2-1}{``Isotropic''\\medium, \( \varepsilon_i \)} & \Block{2-1}{Anisotropic\\medium, \( \hat{\varepsilon} \)\tabularnote{In principal axes. If two are specified, then this is the range of change in the wavelength range.}} & \Block{2-1}{Wavelentgh} & \Block{2-1}{Ref.} \\
        \cmidrule(l{0.2cm}r{0.5cm}){1-3}
        \( \varepsilon_i \)  &  \( \varepsilon_o \) &  \( \varepsilon_e \) & & & & \\
        \midrule
        \( + \) & \( + \) & \( + \) & \Block[t]{}{slightly biaxial,\\\RN{1} in Fig.~\Figref[(a)]{fig:dsw-polariton}} & \Block[t]{}{index-matching liquid\\\( 3.15 \)--\( 3.20 \)} & \Block[t]{}{\ce{KTiOPO4}\\\( (3.10, 3.14, 3.48) \)} & \( 632.8 \)\,nm & \cite{Takayama2009ObservationDyakonov}\\[0.2cm]
        &  &  & \RN{1} in Fig.~\Figref[(a)]{fig:dsw-polariton} & \Block[t]{}{polycarbonate\\\( 2.49 \)} & \Block[t]{}{NLC\tabularnote{nematic liquid crystal} 5CB\\\( (2.38, 2.38, 2.85) \)} & \( 632.8 \)\,nm & \cite{Li2020ControllableSelective}\\[0.2cm]
        &  &  & \RN{1} in Fig.~\Figref[(a)]{fig:dsw-polariton} & \Block[t]{}{\ce{MgF2}\\\( (1.89, 1.89, 1.93) \)} & \Block[t]{}{CSTF\tabularnote{chiral sculptured thin film} \ce{ZnSe} \\ ?} & \( 633 \)\,nm & \cite{Pulsifer2013ObservationDyakonovTamm} \\[0.2cm]
        \( - \) & \( + \) & \( + \) & \RN{2} in Fig.~\Figref[(a)]{fig:spp-types} & \Block[t]{}{\ce{Au}\\\( -11.7 + i 0.73 \)} & \Block[t]{}{\ce{SiO2} film \( 2.15 \)\,nm \\ and NLC\tabularnote{nematic liquid crystal} K15\\\( (2.30, 2.30, 2.95) \)} & \( 632.8 \)\,nm & \cite{Yang1982DeterminationLiquid} \\[0.55cm]
        & & & \RN{2} in Fig.~\Figref[(a)]{fig:spp-types} & \Block[t]{}{\ce{Al} \\ \( -46.40 + i 16.88 \)} & \Block[t]{}{\ce{CaCO3} \\ \( (2.21, 2.21, 2.75) \)} & \( 632.8 \)\,nm & \cite{Simon1995SurfaceElectromagnetic} \\[0.2cm]
        & & & \RN{4} in Fig.~\Figref[(a)]{fig:spp-types} & \Block[t]{}{\ce{Ag}\\\( \approx -8.66 + i 0.61 \)\tabularnote{differs from data in other works \cite{Polyanskiy2024RefractiveindexInfo}}} & \Block[t]{}{\( p \)-6P\tabularnote{\textit{para}-sexiphenyl} nanowires \\ \( \approx(4.6, 4.6, 6.6) \)--\\ \( (3.2, 3.2, 3.9) \)} & \( 495 \)--\( 680 \)\,nm & \cite{Takeichi2011AnisotropicPropagation} \\[0.55cm]
        \( + \) & \( - \) & \( - \) & \Block[t]{}{Not applicable,\\ biaxial} & \Block[t]{}{air \\ \( 1 \)} & \Block[t]{}{nanostructured \ce{Au} film \\ \( (-1.3, -2.5, -3.9) \)--\\\( (-5.0, -7.9, -11.4) \)\tabularnote{without imaginary part}} & \( 540 \)--\( 680 \)\,nm & \cite{Feng2010FormBirefringence} \\[0.55cm]
        &  &  & \RN{1}, \RN{2} in \Figref[(a)]{fig:spp-types-anisotropic-metal} & \Block[t]{}{air} & \Block[t]{}{\ce{Ag} grating \\ (depend on \( \omega \))} & \( 530 \)--\( 700 \)\,nm & \cite{High2015VisiblefrequencyHyperbolic} \\[0.2cm]
        \( + \) & \( - \) & \( + \) & \RN{1}, \RN{2} in Fig.~\Figref[(a)]{fig:hyperbolic-2-polaritons} & \Block[t]{}{air} & \Block[t]{}{deep-trench AZO\tabularnote{\ce{Al} doped \ce{ZnO}} \\ \( (-1, -1, 1.5) \)--\\ \( (-10, -10, 1.5) \)\tabularnote{without imaginary part}} & \( 4 \)--\( 14 \)\,\( \mu \)m & \cite{Takayama2017MidinfraredSurface,Takayama2018ExperimentalObservation} \\[0.55cm]
        & & & \RN{2} in Fig.~\Figref[(a)]{fig:hyperbolic-2-polaritons} & \Block[t]{}{air} & \Block[t]{}{\ce{CaCO3} \\ \( (-3.7, -3.7, 2.3) \)\tabularnote{without imaginary part}} & \( 6.8 \)--\( 7 \)\,\( \mu \)m & \cite{Ma2021GhostHyperbolic} \\[0.2cm]
        \( + \) & \( + \) & \( - \) & \Block[t]{}{slightly biaxial,\\\RN{5} in Fig.~\Figref[(a)]{fig:hyperbolic-1-polaritons}} & \Block[t]{}{air} & \Block[t]{}{nanostructured h\ce{BN}\tabularnote{hexagonal boron nitride} \\ \( (2.1, 3.7, -15.2+i0.6) \)} & \( 7 \)--\( 7.1 \)\,\( \mu \)m & \cite{Li2018InfraredHyperbolic} \\[0.2cm]
        \( + \) & \( \pm \) & \( \pm \) &  Figs.~\Figref[(a)]{fig:spp-types-anisotropic-metal}, \Figref[(a)]{fig:hyperbolic-1-polaritons} & \Block[t]{}{air} & \Block[t]{}{doped hexagonal \ce{ZnO} \\ (depend on \( \omega \))\tabularnote{The authors assumed that the following three terms make the main contribution: static high-frequency \( \varepsilon_\infty \), optical phonons \( \qty(\omega_\text{L}^2 - \omega_\text{T}^2)/\qty(\omega_\text{T}^2 - \omega^2 - i \omega \gamma_\text{ph})\), free carriers \( \omega_p^2 / \qty(\omega^2 - i \gamma_p \omega) \). All terms are assumed to be uniaxially anisotropic.}} & \( 16 \)--\( 25 \)\,\( \mu \)m & \cite{Melnichuk1998SurfacePlasmonphonon} \\[0.2cm]
        & & & Figs.~\Figref[(a)]{fig:spp-types-anisotropic-metal}, \Figref[(a)]{fig:hyperbolic-1-polaritons}, \Figref[(a)]{fig:hyperbolic-2-polaritons} & \Block[t]{}{air} & \Block[t]{}{\ce{Al2O3}\tabularnote{multiple Lorentz oscillators model}} & \( 13 \)--\( 28.5 \)\,\( \mu \)m & \cite{Chu1986DispersionSurface} \\
        & & & Figs.~\Figref[(a)]{fig:spp-types-anisotropic-metal}, \Figref[(a)]{fig:hyperbolic-1-polaritons}, \Figref[(a)]{fig:hyperbolic-2-polaritons} & \Block[t]{}{air} & \Block[t]{}{\ce{TiO2}, \ce{MgF2}\tabularnote{multiple Lorentz oscillators model}} & \( 16 \)--\( 55 \)\,\( \mu \)m & \cite{Bryksin1973SurfaceOscillations} \\
        & & & Figs.~\Figref[(a)]{fig:spp-types-anisotropic-metal}, \Figref[(a)]{fig:hyperbolic-1-polaritons}, \Figref[(a)]{fig:hyperbolic-2-polaritons} & \Block[t]{}{air} & \Block[t]{}{\( \alpha \)-\ce{SiO2}\tabularnote{multiple Lorentz oscillators model}} & \( 8.3 \)--\( 28.5 \)\,\( \mu \)m & \cite{Falge1973DispersionPhononLike,Falge1974DispersionGeneralized} \\
        & & & Figs.~\Figref[(a)]{fig:spp-types-anisotropic-metal}, \Figref[(a)]{fig:hyperbolic-1-polaritons}, \Figref[(a)]{fig:hyperbolic-2-polaritons} & \Block[t]{}{air} & \Block[t]{}{\( \alpha \)-\ce{LiIO3}\tabularnote{multiple Lorentz oscillators model}} & \( 11.7 \)--\( 13.3 \)\,\( \mu \)m & \cite{Puchkovskaya1978CrystalOptics} \\
        \botrule
    \end{NiceTabular}
\end{table*}

\subsection{Summary}
The results of the analysis are summarized in Table~\ref{tab:polariton-classification}.
Each line of the table shows the main features of the surface polariton, based on its isofrequency contour shape, and the conditions for its existence.
The first three columns are self-explanatory.
The fourth column describes what curve an isofrequency contour resembles.
``\( 0 \)--hyperbola'' means that contour looks like a hyperbola with foci on the optic axis.
Similar to a ``\( \pi/2 \)--hyperbola'', but the foci are on the perpendicular to the optic axis.
Finite and infinite arc means whether the wavevector \( q \) of the surface polariton is limited or not at any propagation angle \( \varphi \).
Parameter \( s \) range corresponds to the permissible values of \( s \) in algebraic solution \eqref{eq:exact-solution}--\eqref{eq:exact-solution-angle}.
AED column describes the range of angles \( \varphi \) relative to the optic axis along which surface polariton propagation is possible.
Existence condition column describes restrictions on dielectric permittivities of the media under which the corresponding isofrequency contour shape is implemented.
The last column indicates whether the singular surface polariton \cite{Marchevskii1984SingularElectromagnetic,Fedorov1985SingularWaves,Zhou2019SurfaceplasmonpolaritonWave,Zhou2021SingularExistence} exists and an additional condition if it does.
Recall that the angle at which it propagates is equal to \( \varphi_\text{SSP} \) \eqref{eq:ssp-angle}.

Table~\ref{tab:experiments} shows examples of pairs of materials in which the surface polaritons discussed in this work were experimentally studied.
It mentions only a part of the works listed in the introduction, which seemed the most interesting for further research.
The fourth column indicates which parameter domain \( (\eta, \chi) \) the pair of materials from the fifth and sixth columns falls into.
The fifth and sixth columns indicate the name of the material and the range of dielectric permittivity (or dielectric tensor in principal axis) for the wavelength range in which they were studied.
As can be seen, experimental methods make it possible to study surface polaritons in a wide range of wavelengths.
However, it should be noted that in many works only the most symmetrical directions were considered to simplify further analysis.
More examples of potential materials could be found in the reviews \cite{Borstel1977SurfacePhononpolaritons,Korzeb2015CompendiumNatural,Narimanov2015NaturallyHyperbolic,Takayama2008DyakonovSurface} and in the collection~\cite{Polyanskiy2024RefractiveindexInfo}.

\section{Conclusion}
For a long time it was believed that dispersion equations for surface polaritons in an anisotropic medium can only be solved numerically, with the exception of weakly anisotropic media and highly symmetric propagation directions.
The paper \cite{Kroytor2021InvestigationExistence} was the first to show the existence of an exact solution using computer algebra system, but the analysis was done only for Dyakonov surface waves.
In this work we extend this result to all types of surface polaritons propagating at the boundary of isotropic and uniaxial medium in the case of the optic axis parallel to the boundary.
The complete solution in algebraic form \eqref{eq:exact-solution}--\eqref{eq:exact-solution-angle} is written for arbitrary ratios between the dielectric permittivities of media, neglecting losses.
In particular, the obtained solution may be useful in analyzing the properties of surface polaritons in metamaterials, where the weak anisotropy approximation is not applicable.
Some cases are considered in this work that have not been discussed in detail previously in the literature.
The existence of surface polaritons at the interface of a metal-like medium and a Type I hyperbolic medium is predicted.
Perhaps elliptic surface polaritons at the boundary of an anisotropic metal were considered for the first time.

The dependence of the wavevector of the surface polariton on the propagation angle relative to the optic axis have been analyzed.
As a result, possible shapes of the isofrequency contour have been obtained depending on the relations between the dielectric permittivities.
In general, a surface polariton is shown to be one of three types, depending on the shape of the contour: \textit{elliptic-like}, \textit{hyperbolic-like} and \textit{arc-like} (Dyakonov surface waves).
The contour shape changes with frequency if there is a frequency dispersion of the permittivity of media.
The results obtained can be used in the analysis of any types of surface polaritons, regardless of their origin (phonons, plasmons, etc.), at the boundary of a uniaxial and isotropic medium, described only by the dielectric permittivity tensor.

\begin{acknowledgments}
    The work was supported by Theoretical Physics and Mathematics Advancement Foundation ``BASIS'' and by Clover Program: Joint Research Projects of Skoltech, MIPT, and ITMO.
    The author thanks N.\,S.~Averkiev and N.\,A.~Gippius for fruitful discussions.
\end{acknowledgments}

\appendix
\section{Field components and dispersion equation\label{sec:fields-expressions}}
Following the statement of the problem we are looking for a wave propagating along the boundary between two media.
If we exclude purely electric waves (\( \vb{H} = 0 \)) then in any anisotropic linear medium there are two eigenwaves that differ in polarization.
Typically, in media without spatial dispersion, longitudinal purely electric waves exist only at certain frequencies.
We do not consider these cases, assuming that there are no longitudinal waves at frequency \( \omega \).
Thus, the general solution is the sum of two eigenwaves in each medium.
In an anisotropic medium we have to choose an ordinary and an extraordinary wave with the exception of directions called singular axes \cite{Voigt1902BehaviourPleochroitic,Fedorov1976TeorijaGirotropii,Marchevskii1984SingularElectromagnetic,Fedorov1985SingularWaves}.
Let us choose an eigenvector for the ordinary wave as
\begin{equation}\label{eq:ordinary-waves-polarization}
    \vb{E}_o = \begin{pmatrix} q_y \\ i \kappa_o \\ 0\end{pmatrix}\qcomma \vb{H}_o = k_0^{-1} \begin{pmatrix} -i \kappa_o q_z \\ q_y q_z \\ -(\varepsilon_o k_0^2 - q_z^2)\end{pmatrix},
\end{equation}
and for the extraordinary wave as
\begin{equation}\label{eq:extraordinary-waves-polarization}
    \vb{E}_e = \begin{pmatrix} -i\kappa_e q_z \\ q_y q_z \\ -(\varepsilon_o k_0^2 - q_z^2)\end{pmatrix}\qcomma \vb{H}_e = \varepsilon_o k_0 \begin{pmatrix} -q_y \\ -i \kappa_e \\ 0\end{pmatrix},
\end{equation}
where \( (q_y, q_z) \) is the in-plane wavevector, \( \kappa_o \) and \( \kappa_e \) are decrements of field decay from the boundary, and \( k_0 = \omega / c \) is the wavevector in vacuum.
The expressions are obtained from solving the Fresnel equation system \cite{Born2019PrinciplesOptics}.

The form of eigenvectors \eqref{eq:ordinary-waves-polarization} and \eqref{eq:extraordinary-waves-polarization} is especially useful.
If we put \( \varepsilon_o = \varepsilon_e = \varepsilon_i \) as in an isotropic medium, then still form a complete basis, like TE- and TM-waves.
The transition from one basis to another is given \( \vb{E}_\text{TE} \propto (-i \kappa_i q_z \vb{E}_o - q_y \vb{E}_e) \) and \( \vb{H}_\text{TM} \propto (-i \kappa_i q_z \varepsilon_i^{-1} \vb{H}_e + q_y k_0^2 \vb{H}_o) \), the other components can be derived from Maxwell's equations.
The relations between wavevector components are given by \eqref{eq:dispersion-equation-isotropic}--\eqref{eq:dispersion-equation-extraordinary}.

Using \eqref{eq:ordinary-waves-polarization} and \eqref{eq:extraordinary-waves-polarization}, the field distributions for a surface wave are written in an anisotropic medium \( x < 0 \) as
\begin{equation*}\label{eq:fields-uniaxial-medium}
    \begin{aligned}
        &\vb{E}_{x<0} = \qty\big( a_1 \vb{E}_o e^{\kappa_o x} + a_2 \vb{E}_e e^{\kappa_e x} )\, e^{i q_y y + i q_z z - i \omega t}, \\
        &\vb{H}_{x<0} = \qty\big( a_1 \vb{H}_o e^{\kappa_o x} + a_2 \vb{H}_e e^{\kappa_e x} )\, e^{i q_y y + i q_z z - i \omega t},
    \end{aligned}
\end{equation*}
and in an isotropic medium \( x > 0 \) as
\begin{equation*}\label{eq:fields-isotropic-medium}
    \begin{aligned}
        &\vb{E}_{x>0} = \qty\big( b_1 \vb{E}_o^{(i)} + b_2 \vb{E}_e^{(i)} )\, e^{-\kappa_i x}\, e^{i q_y y + i q_z z - i \omega t},\\
        &\vb{H}_{x>0} = \qty\big( b_1 \vb{H}_o^{(i)} + b_2 \vb{H}_e^{(i)} )\, e^{-\kappa_i x}\, e^{i q_y y + i q_z z - i \omega t},
    \end{aligned}
\end{equation*}
where superscript \( {(i)} \) means that we put \( \varepsilon_o = \varepsilon_e = \varepsilon_i \) in \eqref{eq:ordinary-waves-polarization} and \eqref{eq:extraordinary-waves-polarization}.

In the singular case we have \( q_z^2 = \varepsilon_o k_0^2 \) and accordingly \( \kappa_o^2 = \kappa_e^2 = q_y^2 \) from \eqref{eq:dispersion-equation-ordinary}--\eqref{eq:dispersion-equation-extraordinary}.
Where it immediately follows \( \vb{E}_e \propto \vb{E}_o \) and \( \vb{H}_e \propto \vb{H}_o\) means that the chosen basis is not complete.
In this case we write the general solution \cite{Marchevskii1984SingularElectromagnetic,Fedorov1985SingularWaves}, correctly calculating \( \vb{H} \), as
\begin{equation*}
    \begin{aligned}
        &\vb{E}_{x<0}^{(\text{SSP})} = \begin{pmatrix} q_y \\ i q_y \\ 0\end{pmatrix}\qty\big( a_1 + a_2 x )\,e^{q_y x}\,e^{i q_y y + i q_z z - i \omega t}, \\
        &\vb{H}_{x<0}^{(\text{SSP})} = \begin{pmatrix} -i q_y q_z (a_1 + a_2 x)\\ q_y q_z (a_1 + a_2 x) \\ a_2\end{pmatrix} e^{q_y x}\,e^{i q_y y + i q_z z - i \omega t}
    \end{aligned}
\end{equation*}
Even though the singular solution has a different form, this does not affect the dispersion equation.
The exact solution for singular surface polariton is given in \cite{Marchevskii1984SingularElectromagnetic,Lakhtakia2020ElectromagneticSurface}.

Now using the boundary conditions of continuity of the tangential components \( E_{y,z} \) and \( H_{y,z} \), we obtain the linear system
\begin{equation}\label{eq:surface-polariton-linear-system}
    \underbrace{\begin{pmatrix}
        i \kappa_o & q_y q_z & i \kappa_i & - q_y q_z \\
        q_y q_z & -i \kappa_e \varepsilon_o k_0^2 & - q_y q_z & - i \kappa_i \varepsilon_i k_0^2 \\
        0 & g_o & 0 & -g_i \\
        g_o & 0 & -g_i & 0
    \end{pmatrix}}_{M} \begin{pmatrix}
        a_1\\a_2\\b_1\\b_2
    \end{pmatrix} = 0,
\end{equation}
where \( g_o = q_z^2 - \varepsilon_o k_0^2 \) and \( g_i = q_z^2 - \varepsilon_i k_0^2 \).
Nontrivial solution exist if \( \det M = 0 \).
The matrix \( M \) can be viewed as a block matrix
\begin{equation*}
    M = \begin{pmatrix}A & B \\C & D\end{pmatrix}
\end{equation*}
with \( 2\!\times\!2 \) blocks.
It is clear that \( CD = DC \).
Then \( \det M \) can be easily calculated as
\begin{multline*}
    \det M = \det (AD - BC) =\\
    = \det \begin{pmatrix}
        q_y q_z (g_o - g_i) & -i(\kappa_i g_o + \kappa_o g_i) \\
        i k_0^2 (\varepsilon_i \kappa_i g_o + \varepsilon_o \kappa_e g_i) & q_y q_z (g_o - g_i)
    \end{pmatrix}.
\end{multline*}
Calculating the last determinant yields dispersion equation \eqref{eq:dispersion-equation}.
The ratios of the contributions of different polarizations can be calculated from \eqref{eq:surface-polariton-linear-system}, putting, for example, \( a_2 = 1 \).

\section{Dispersion equation simplification\label{sec:angle-elimination}}
The dispersion equation \eqref{eq:dispersion-equation} has two remarkable properties that will be useful.
It is easy to see that substituting \( n_z^2 = \varepsilon_i \) turns it into an identity.
However, such a substitution does not help to obtain a solution for \( n_y \).
The chosen polarizations of waves \eqref{eq:ordinary-waves-polarization}--\eqref{eq:extraordinary-waves-polarization} in the isotropic medium are parallel in this case.
The wave with \( E_z \neq 0 \) is lost.
Despite this, the simplified dispersion equation obtained below allows us to obtain \( n_y \) and accordingly the surface polariton wavevector \( q_s \).
Although it will not be possible to determine all field components.
Similarly, if we put \( n_z^2 = \varepsilon_o \) then \eqref{eq:dispersion-equation} becomes \( \kappa_i \kappa_e = n_y^2 \).
We can prove that this is an identity if both \( \kappa_i, \kappa_e > 0 \).

Let us explicitly take out \( \qty(n_z^2 - \varepsilon_i)\qty(n_z^2 - \varepsilon_o) \) in \eqref{eq:dispersion-equation}.
Let \( f_1 = n_z^2 - \varepsilon_i \) and \( f_2 = n_z^2 - \varepsilon_o \).
Then \eqref{eq:dispersion-equation} becomes
\begin{multline}\label{eq:dispersion-equation-f}
    \qty\big(\kappa_i f_2 + \kappa_o f_1) \qty\big(\kappa_i \varepsilon_i f_2 + \kappa_e \varepsilon_o f_1) - \\
    - n_y^2 n_z^2 \qty(\varepsilon_i - \varepsilon_o)^2 = 0.
\end{multline}

Let us transform the last term
\begin{multline*}
    n_y^2 n_z^2 \qty(\varepsilon_i - \varepsilon_o)^2 = n_y^2 n_z^2 \qty(f_2 - f_1)^2 = \\
    = n_y^2 \qty\Big[\qty(f_1 + \varepsilon_i) \qty(f_2^2 - f_1 f_2) + \qty(f_2 + \varepsilon_o) \qty(- f_1 f_2 + f_1^2)] = \\
    = - n_y^2 (\varepsilon_i + \varepsilon_o) f_1 f_2 + n_y^2 \varepsilon_i f_2^2 + n_y^2 \varepsilon_o f_1^2.
\end{multline*}
Combining terms \( \propto f_1^2 \) and \( \propto f_2^2 \) with the first term in \eqref{eq:dispersion-equation-f} yields
\begin{gather*}
    \qty(\kappa_i^2 - n_y^2) \varepsilon_i f_2^2 + \varepsilon_o \qty(\kappa_o \kappa_e - n_y^2) f_1^2 + \mathcal{A} f_1 f_2 = 0,\\
    \mathcal{A} = \kappa_i \qty(\varepsilon_i \kappa_o + \varepsilon_o \kappa_e) + n_y^2 \qty(\varepsilon_i + \varepsilon_o).
\end{gather*}
It follows from \eqref{eq:dispersion-equation-isotropic} that
\begin{equation*}
    \qty(\kappa_i^2 - n_y^2) \varepsilon_i = \varepsilon_i f_1.
\end{equation*}
Combining \eqref{eq:dispersion-equation-ordinary} and \eqref{eq:dispersion-equation-extraordinary} together yields
\begin{multline*}
    \varepsilon_o \qty(\kappa_o \kappa_e - n_y^2) = \kappa_o\varepsilon_o\qty(\kappa_e - \kappa_o) + \varepsilon_o f_2 =\\
    = \frac{\varepsilon_e - \varepsilon_o}{\kappa_e + \kappa_o} \kappa_o f_2 + \varepsilon_o f_2
\end{multline*}

Using the obtained relations, Eq.~\eqref{eq:dispersion-equation-f} can be rewritten as
\begin{multline}\label{eq:dispersion-equation-f1f2}
    \biggl(\varepsilon_i \qty[\kappa_i \kappa_o + n_y^2 + f_2] + \varepsilon_o \qty[\kappa_i \kappa_e + n_y^2 + f_1] + \\
    + \frac{\varepsilon_e - \varepsilon_o}{\kappa_e + \kappa_o} \kappa_o f_1\biggr) f_1 f_2 = 0.
\end{multline}
The first square bracket is \( \kappa_i \kappa_o + n_y^2 + f_2 = \kappa_o (\kappa_i + \kappa_o)\).
The second square bracket is \( \kappa_i \kappa_e + n_y^2 + f_1 = \kappa_i (\kappa_e + \kappa_i) \).
It follows from \eqref{eq:dispersion-equation-ordinary}--\eqref{eq:dispersion-equation-extraordinary} that \( \varepsilon_o \kappa_e^2 - \varepsilon_e k_o^2 = (\varepsilon_o - \varepsilon_e) n_y^2 \).
Using \( f_1 = \kappa_i^2 - n_y^2 \) and rearranging the remaining terms we obtain
\begin{multline}\label{eq:dispersion-remaining-terms}
    \varepsilon_o \kappa_i \qty( \kappa_e + \kappa_i) \qty(\kappa_e + \kappa_o) + (\varepsilon_e - \varepsilon_o) \kappa_o \kappa_i^2\ + \\
    + \kappa_o \qty(\varepsilon_o \kappa_e^2 - \varepsilon_e \kappa_o^2) = \\
    = \qty\Big[\varepsilon_o \kappa_e \qty(\kappa_e + \kappa_i) - \varepsilon_e \kappa_o \qty(\kappa_o - \kappa_i)]\qty(\kappa_o + \kappa_i).
\end{multline}

After collecting all pieces together we obtain that Eq.~\eqref{eq:dispersion-equation-f} is equivalent to
\begin{multline}\label{eq:dispersion-equation-simpliest}
    \qty\Big[\qty(\varepsilon_i \kappa_o + \varepsilon_o \kappa_e)\qty(\kappa_e + \kappa_i) + (\varepsilon_i - \varepsilon_e)\kappa_o\qty(\kappa_o - \kappa_i)] \times \\
    \times \frac{\kappa_i + \kappa_o}{\kappa_e + \kappa_o} f_1 f_2 = 0.
\end{multline}
Remembering

If we multiply the expression in square brackets by \( {(\kappa_o + \kappa_i)} \), then the dispersion equation takes form~\cite[Eq.~(9) in][]{Dyakonov1988NewType}
\begin{equation}
    (\kappa_i + \kappa_e)(\kappa_i + \kappa_o)(\varepsilon_i\kappa_o + \varepsilon_o\kappa_e) = (\varepsilon_e - \varepsilon_i)(\varepsilon_i - \varepsilon_o) \kappa_o.
\end{equation}
On the other hand, if we leave \( n_y^2 \) and substitute \( \varepsilon_o \kappa_e^2 \) in terms of \( \kappa_o^2 \) then Eq.~\eqref{eq:dispersion-remaining-terms} has the form~\citep[Eq.~(5) in][]{Alshits2002DispersionlessSurface}
\begin{equation}\label{eq:dispersion-dsw-alshits}
    \varepsilon_i \kappa_o^2 + \kappa_e (\varepsilon_i \kappa_o + \varepsilon_o \kappa_i) + \varepsilon_e \kappa_i \kappa_o = (\varepsilon_e - \varepsilon_o) n_y^2.
\end{equation}

\section{Algebraic solution derivation\label{sec:dispersion-solution}}
Let us start from dispersion equation~\eqref{eq:dispersion-equation-simpliest} considering that \( f_1 \neq 0 \) and \( f_2 \neq 0 \).
\begin{equation}\label{eq:dispersion-quadratic-form}
    \qty(\varepsilon_i \kappa_o + \varepsilon_o \kappa_e)\qty(\kappa_e + \kappa_i) + (\varepsilon_i - \varepsilon_e)\kappa_o\qty(\kappa_o - \kappa_i) = 0.
\end{equation}
One can notice that \eqref{eq:dispersion-quadratic-form} is a quadratic form of variable \( \kappa_i, \kappa_o, \kappa_e \), and it only has linear \( \kappa_i \) terms.
We can isolate \( \kappa_i \) terms
\begin{equation*}
    \kappa_i \qty(\varepsilon_e \kappa_o + \varepsilon_o \kappa_e)
    = \varepsilon_i \kappa_o \qty(\kappa_o + \kappa_e) + \qty(\varepsilon_o \kappa_e^2 - \varepsilon_e \kappa_o^2).
\end{equation*}
If we want to square it then both sides must have the same sign.
The set of  equation for squares then additional roots may appear.
On the other hand, we can express \( \kappa_i^2 \) through \( \kappa_o^2 \).
The resulting equation includes only the 2 and 4 powers of \( \kappa_o \) and \( \kappa_e \) which are combined into two homogeneous polynomials \( P_2(\kappa_o, \kappa_e) \) and \( P_4(\kappa_o, \kappa_e) \) of the 2 and 4 degree respectively~\cite{Kroytor2021InvestigationExistence}
\begin{equation}\label{eq:polynomials-sum}
    P_2(\kappa_o, \kappa_e) + P_4(\kappa_o, \kappa_e) = 0,
\end{equation}
where
\begin{gather*}
    P_2(\kappa_o, \kappa_e) = \qty(\varepsilon_i - \varepsilon_o)\qty(\varepsilon_e \kappa_o + \varepsilon_o \kappa_e)^2, \\
    \begin{multlined}
        P_4(\kappa_o, \kappa_e) = \qty(\kappa_o + \kappa_e)\qty(\varepsilon_i \kappa_o + \varepsilon_o \kappa_e) \times \\ \times \qty\Big[(\varepsilon_i - \varepsilon_o) \kappa_o \kappa_e + (\varepsilon_i - 2 \varepsilon_e) \kappa_o^2 + \varepsilon_o \kappa_e^2].
    \end{multlined}
\end{gather*}
By definition, a polynomial \( P_n(x, y) \) is called homogeneous polynomial of the degree \( n \) if \( {P_n(\lambda x, \lambda y) = \lambda^n P_n(x, y)} \) for any \( \lambda \).

Now, Eq.~\eqref{eq:polynomials-sum} can be easily solved for \( \kappa_e \) or \( \kappa_o \), e.g. by introducing the parameter \( s = \kappa_e / \kappa_o \).
In our case it should be \( s > 0 \) because \( \kappa_e, \kappa_o > 0 \) for a surface polariton.
Substituting \( \kappa_e \) in \eqref{eq:polynomials-sum} and canceling by \( \kappa_o^2 \), we obtain equation
\begin{equation*}
    P_2(1, s) + \kappa_o^2 P_4(1, s) = 0.
\end{equation*}
Let put \( P_2(s) \equiv P_2(1, s) \) and \( P_4(s) \equiv P_4(1, s) \).
Solving equation for \( \kappa_o \) we obtain \eqref{eq:exact-solution}.

Introducing inverse relation \( \bar{s} = \kappa_o / \kappa_e = s^{-1} \) makes more sense in some cases.
If \( s \to 0 \) then \( \kappa_e \to 0 \) too, regardless \( \kappa_o \) and \( \kappa_i \).
This means that the surface polariton is becoming less localized due to the long decay length for the extraordinary component \( \propto 1 / \kappa_e \).
On the other hand, if \( \bar{s} \to 0 \) then \( \kappa_o \to 0 \), and polariton is becoming less localized due to increasing decay length for the ordinary component \( \propto 1 / \kappa_o \).

\section{Directions of high symmetry\label{sec:high-symmetry-directions}}
For highly symmetric directions, along and perpendicular to the optic axis, dispersion equation \eqref{eq:dispersion-equation} is significantly simplified.
This is explained by the fact that along these directions the ordinary and extraordinary waves are not mixed by the boundary conditions.

If a surface polariton propagates along the optic axis then \( n_y = 0 \) and accordingly \( \kappa_e^2 = (\varepsilon_e / \varepsilon_o) \kappa_o^2 \) from \eqref{eq:dispersion-equation-ordinary}--\eqref{eq:dispersion-equation-extraordinary}.
Thus, the second bracket in dispersion equation \eqref{eq:dispersion-equation} should vanish.
It brings us to the equation
\begin{equation}\label{eq:dispersion-along}
    \frac{\kappa_e}{\varepsilon_e} + \frac{\kappa_i}{\varepsilon_i} = 0,
\end{equation}
which is similar to well-known SPP dispersion equation.
The solution is
\begin{equation}\label{eq:solution-along}
    n_z^2 = \frac{\varepsilon_i \varepsilon_o\qty(\varepsilon_i - \varepsilon_e)}{\varepsilon_i^2 - \varepsilon_o \varepsilon_e} = \frac{\varepsilon_o \chi (1 + \eta + \chi)}{\chi^2 - 1 - \eta}
\end{equation}
and exists in two cases: \( \varepsilon_i < 0 \) and \( \varepsilon_o, \varepsilon_e > 0 \); \( \varepsilon_i > 0 \) and \( \varepsilon_o, \varepsilon_e < 0 \) \cite{Agranovich1975CrystalOptics,Burstein1974SurfacePolaritons,Agranovich1982SurfacePolaritons}.
Also, we need to require \( n_z \) to be real.
If \( \varepsilon_o > 0 \) then the additional condition is \( \chi > \sqrt{1 + \eta} \) in our notation.
This corresponds to domains \RN{1}, \RN{4} and \RN{5} in Fig.~\Figref[(a)]{fig:spp-types}.
In the case \( \varepsilon_o < 0 \) the condition \( 0 < \chi < \sqrt{1 + \eta} \) should be satisfied.
The corresponding domains are \RN{1} and \RN{2} (Fig.~\Figref[(a)]{fig:spp-types-anisotropic-metal}).
It can be seen from \eqref{eq:extraordinary-waves-polarization} and \eqref{eq:surface-polariton-linear-system} that only the ``extraordinary'' wave in anisotropic medium and TM-wave in isotropic medium remain.

It may seem from the right side of Eq.~\eqref{eq:solution-along} that the solution is also possible for \( \varepsilon_e < 0 \) and \( \varepsilon_o > 0 \).
This is incorrect.
Any deviation from the optic axis, no matter how small, leads to the mixing of an ordinary wave.
The wavevector \( q \) of the surface polariton cannot change significantly.
The attenuation decrement \( \kappa_o \) of ordinary wave (Eq.~\eqref{eq:dispersion-equation-ordinary}) is imaginary for such \( q \).
It means that the ordinary wave component of a polariton is not localized at the interface.
Thus this solution does not correspond to a well defined surface polariton.
Such type of solution is called a virtual surface polariton in \cite{Burstein1974SurfacePolaritons,Agranovich1982SurfacePolaritons}.

If the direction of propagation is perpendicular to the optic axis then \( n_z = 0 \).
The only way to satisfy Eq.~\eqref{eq:dispersion-equation} is to make the first bracket equal to zero.
This yields a dispersion equation similar to \eqref{eq:dispersion-along}
\begin{equation*}
    \frac{\kappa_o}{\varepsilon_o} + \frac{\kappa_i}{\varepsilon_i} = 0,
\end{equation*}
and its solution is
\begin{equation*}
    n_y^2 = \frac{\varepsilon_o \varepsilon_i}{\varepsilon_o + \varepsilon_i}.
\end{equation*}
It is the well-known solution for SPP wavevector at the boundary of two isotropic media with dielectric permittivities \( \varepsilon_o \) and \( \varepsilon_i \).
To the usual SPP existence conditions \( \varepsilon_o + \varepsilon_i < 0 \) and \( \varepsilon_o \varepsilon_i < 0 \), we need to add \( n_y^2 > \varepsilon_e \) for the same reason as in the previous case.
In our notation they can be expressed as follows
\begin{align*}
    &\varepsilon_i < 0:\quad \begin{cases}
        \chi > 1 &\qif \eta \leqslant 0,\\
        1 < \chi < 1 + \eta^{-1} &\qif \eta > 0,
    \end{cases}\\
    &\varepsilon_i > 0:\quad \begin{cases}
        \chi < 1 &\qif \eta \geqslant -1,\\
        1 + \eta^{-1} < \chi < 1 &\qif \eta < -1.
    \end{cases}
\end{align*}
The corresponding domains for the first case are \RN{3} and \RN{4} in Fig.~\Figref[(a)]{fig:spp-types}, \RN{1} and \RN{2} in Fig.~\Figref[(a)]{fig:hyperbolic-1-polaritons}.
For the second case are \RN{2} and \RN{3} in Fig.~\Figref[(a)]{fig:spp-types-anisotropic-metal}, \RN{1} in Fig.~\Figref[(a)]{fig:hyperbolic-2-polaritons}.
%The ordinary wave remain in anisotropic medium for this direction \eqref{eq:extraordinary-waves-polarization} and \eqref{eq:surface-polariton-linear-system}.

\bibliography{anisotropic-surface-polaritons.bib}

\end{document}